\documentclass[12pt]{article}

\usepackage{amsmath}
\usepackage{amssymb}
\usepackage{epsf}

\title{%
\Large
\textbf{Detonations and deflagrations in cosmological \\
phase transitions}}
\vspace{1cm}
\author{ \large Ariel M\'egevand\footnote{Member of CONICET,
Argentina. E-mail address: megevand@mdp.edu.ar}~ and Alejandro D.
S\'anchez\footnote{Member of CONICET, Argentina. E-mail address:
sanchez@mdp.edu.ar} \\[0.5cm]
\normalsize \emph{IFIMAR (CONICET-UNMdP)}, \\
\normalsize \emph{Departamento de F\'{\i}sica, Facultad de Ciencias
Exactas y Naturales,
UNMdP,} \\ \normalsize \emph{De\'an Funes 3350, (7600) Mar del Plata,
Argentina}
\date{}
}

\begin{document}
\maketitle

\begin{abstract}
We study the steady state motion of bubble walls in
cosmological phase transitions. Taking into account the
boundary and continuity conditions for the fluid variables, we
calculate numerically the wall velocity as a function of the
nucleation temperature, the latent heat, and a friction
parameter. We determine regions in the space of these
parameters in which detonations and/or deflagrations are
allowed. In order to apply the results to a physical case, we
calculate these quantities in a specific model, which consists
of an extension of the Standard Model with singlet scalar
fields. We also obtain analytic approximations for the wall
velocity, both in the case of deflagrations and of detonations.
\end{abstract}

\section{Introduction} \label{intro}

Phase transitions of the universe may give rise to a variety of
cosmological relics, such as the baryon asymmetry of the
universe \cite{ckn93}, cosmic magnetic fields \cite{gr01},
topological defects \cite{vs94}, inhomogeneities \cite{h95},
and gravitational waves \cite{kkt94,gw,m08}. To be observable,
several of these relics depend on the strength of the phase
transition. In a first-order phase transition, bubbles of the
stable phase nucleate and grow inside the supercooled phase.
The expansion of bubbles provides a departure from thermal
equilibrium, which is generally required to generate
cosmological remnants. The mechanisms of relic generation
depend in general on the motion of the bubble walls (either
because they are based on charge transport near the bubble
walls, on the collisions of the walls, or on the turbulence
they produce). As a consequence, an important parameter in the
generating mechanisms is the velocity of bubble expansion. For
some relics, higher velocities give a stronger effect. This is
the case, e.g., of gravitational waves. In other cases, on the
contrary, the amplitude or abundance of the relic peaks at some
value of the bubble wall velocity. This happens for instance in
the case of electroweak baryogenesis.

A first-order phase transition requires the system to have two
phases with different equation of state (EOS), which coexist in
some temperature range. Thus, the high-temperature phase has a
pressure $p_+(T)$, and the low-temperature phase has a pressure
$p_-(T)$. At the critical temperature $T_c$, the two phases
have the same pressure but different entropy and energy. The
entropy density is given by $s(T)=p'(T)$, where a prime denotes
derivation with respect to $T$, and the energy density is given
 by $\rho(T)=Ts(T)-p(T)$. The latent heat $L$ is defined as
the difference between the energy densities of the two phases
at $T=T_c$. Therefore, we have $p_+(T_c)=p_-(T_c)$ and
$L=T_c\left(p'_+(T_c)-p'_-(T_c)\right)$. Furthermore, all these
quantities can be derived from the free energy density
$\mathcal{F}$ by the relation $p(T)=-\mathcal{F}(T)$.

In a finite temperature field theory, we have in general a scalar
field $\phi$, the Higgs field, which plays the role of an order
parameter for the phase transition. We shall assume this is the case,
although most of our conclusions will be independent of this
assumption. The free energy density is the finite temperature
effective potential, which is a function of $\phi$ and $T$. In some
range of temperatures, the free energy has two minima separated by  a
barrier. The value of $\phi$ in each minimum defines the two phases.
We shall assume that the high-temperature minimum is $\phi=0$, and
the low temperature minimum has a value $\phi_m(T)\neq 0$, which is
the case for a symmetry breaking phase transition. A phase transition
with a large value of the order parameter, $\phi_m(T)>T$ is usually
called a strongly first-order phase transition. Thus, we have
$\mathcal{F}_+(T)=\mathcal{F}(0,T)$ and
$\mathcal{F}_-(T)=\mathcal{F}(\phi_m(T),T)$. Below the critical
temperature, the transition from $\phi=0$ to $\phi=\phi_m(T)$ occurs
via the nucleation of bubbles \cite{c77,nucl}. A bubble is a
configuration $\phi(x)$ of the field. We shall make the usual
assumption that this configuration is a thin-walled sphere.  A
nucleated bubble grows because $p_-(T)>p_+(T)$ for $T<T_c$.

The propagation of bubble walls in cosmological phase transitions has
been extensively investigated
\cite{ikkl94,dlhll92,lmt92,k92,mp95,mt97,m00,js01,s82,gkkm84,k85,%
eikr92,l94,kl95,mp89,ahm00}. In general, the stronger the phase
transition, the larger the amount of supercooling. Therefore,
one expects that the velocity of bubble walls will be larger
for stronger phase transitions. Indeed, the pressure difference
which pushes the wall of a bubble, $\Delta p(T)
=p_-(T)-p_+(T)$, grows as the temperature decreases.  On the
other hand, the velocity depends also on the friction of the
wall with the medium and the latent heat that is released at
the phase transition front. The friction is related to the
\emph{microphysics}, i.e., to the interactions of particles of
the medium with the Higgs field in the configuration of the
wall, whereas the latent heat is involved in the
\emph{hydrodynamics}, i.e., the bulk motions and reheating of
the fluid near the wall.

According to hydrodynamics, there are two kinds of steady state
solutions for the propagation of the wall, namely, detonations
and deflagrations. Detonations are supersonic while
deflagrations are generally subsonic. Each solution has
different boundary conditions, and both may appear in a given
model and may even coexist in some range of parameters. A
numerical calculation of the evolution of the field $\phi(x,t)$
\cite{ikkl94}, suggests that the bubbles grow typically as
subsonic deflagrations, although, in some regions of parameter
space, the solutions switch from deflagrations to detonations,
as the friction is decreased. In view of the cosmological
consequences of the phase transition, it is important to
determine the range of parameters where each solution may
exist. Hydrodynamic considerations give general constraints on
the allowed regions of thermodynamic parameters, but the
existence of deflagrations and detonations depends strongly on
the value of the friction.

The literature on the subject of the bubble wall velocity can
be roughly divided in two groups: those papers which calculate
the friction for a given model, assuming nonrelativistic
deflagrations and ignoring hydrodynamics
\cite{dlhll92,lmt92,k92,mp95,mt97,m00,js01}, and those which
consider hydrodynamics, but they either disregard microphysics
or include the friction as a free parameter
\cite{ikkl94,s82,gkkm84,k85,eikr92,l94,kl95,mp89}. As a
consequence, the analytical approximations that are most
commonly used in applications, were derived ignoring either
microphysics or hydrodynamics. Indeed, in the context of
baryogenesis the wall velocity is assumed to be of the form
$v_w\approx\Delta p(T)/\eta$, where $\eta$ is a friction
coefficient. This is a nonrelativistic approximation (thus
appropriate for the case of deflagrations) which neglects
hydrodynamics. In the context of gravitational waves, on the
contrary, the wall is assumed to propagate as a detonation, and
the so called Chapman-Jouguet hypothesis is further assumed,
leading to a simple expression for the velocity, which depends
only on the ratio between the latent heat and the thermal
energy density of the plasma.

The aim of this paper is to calculate the wall velocity in both
cases, detonations and deflagrations, taking the friction into
account. This will allow us to investigate under which
conditions the various hydrodynamic solutions can exist, and to
explore how these conditions are attained in a physical model.
For these purposes, we shall first consider the equations for
the discontinuity of the fluid variables across the wall,
together with a simple parametrization of the friction force
(which consists in introducing a damping term in the equation
for the field $\phi$) and a simple model (the bag equation of
state) for the phase transition. These simplifications permit
to write down a complete set of equations, which can be solved
for the wall velocity as a function of three parameters,
namely, $\alpha_c=L/(4\tilde{\rho}_+(T_c))$, where $L$ is the
latent heat and $\tilde{\rho}_+(T)$ is the thermal energy
density of the high-temperature phase,
$\alpha_N=L/(4\tilde{\rho}_+(T_N))$, where $T_N<T_c$ is the
temperature at which bubbles nucleate, and $\eta/L$, where
$\eta$ is the friction coefficient.

Solving numerically these equations, we shall determine the
regions in the space of the aforementioned parameters for which
deflagrations and detonations may or may not exist. In order to
apply the results to a physical model, we choose as an example
the electroweak phase transition in an extension of the
Standard Model (SM) with scalar singlets, which provides a
simple variation of the strength of the phase transition as a
function of the fundamental parameters. For this specific
model, we compute the quantities $\alpha_c$, $\alpha_N$, and
$\eta/L$ and we calculate the wall velocity.  To estimate the
value of the parameter $\eta$, we shall relate our
parametrization of the friction to the microphysics
calculations that exist in the literature. For the nucleation
temperature $T_N$, we solve numerically the bounce action to
obtain the nucleation rate. We shall also find analytical
approximations for both propagation modes. In particular, we
give a simple approximation for the detonation solution, which
depends on the friction.

The paper is organized as follows. In section \ref{eqs} we
write down the equations for the wall velocity, the fluid
variables, and the friction for both propagation modes. In
section \ref{nume} we show the numerical results for the wall
velocity as a function of the relevant parameters, and we
determine regions in parameter space where each kind of
solution may or may not exist. In section \ref{model} we
consider a specific model, for which we compute the values of
the latent heat $L$, the friction $\eta$, and the nucleation
temperature $T_N$, and we calculate the detonation and
deflagration velocities (when they exist) as functions of the
fundamental parameters of the model. Finally, in section
\ref{aprox} we consider analytical approximations for the wall
velocity. Our conclusions are summarized in section
\ref{conclu}.

\section{The wall velocity \label{eqs}}

To take into account hydrodynamics and microphysics in the
calculation of the wall velocity, three basic ingredients are
needed, namely, the continuity conditions that relate the fluid
variables across the wall discontinuity, an equation of state,
which relates the various fluid variables on each side of the
phase transition front, and an equation for the effect of
friction on the motion of the wall.

\subsection{Hydrodynamics}

\begin{figure}[hbt]
\centering \epsfxsize=10cm
\leavevmode \epsfbox{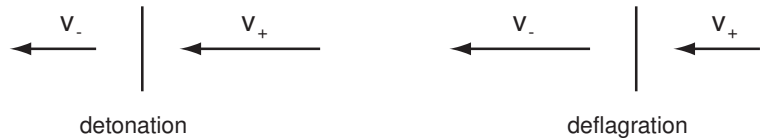}
\caption{The velocities on each side of the wall, in the reference
frame of the wall.}
\label{detodefla}
\end{figure}

In the rest frame of the phase-transition front, we have an incoming
flow in the high-temperature phase, with velocity $v_{+}$, and an
outgoing flow in the low-temperature phase, with velocity $v_{-}$
(see Fig. \ref{detodefla}). The motion of the wall causes the
temperatures $T_{+}$ and $T_{-}$ on each side to be different. The
continuity conditions for energy and momentum fluxes give two
equations \cite{landau},
\begin{equation}
w_{+}\gamma _{+}^{2}v_{+}=w_{-}\gamma _{-}^{2}v_{-},\ w_{+}\gamma
_{+}^{2}v_{+}^{2}+p_{+}=w_{-}\gamma _{-}^{2}v_{-}^{2}+p_{-},
\label{eqlandau}
\end{equation}%
where $\gamma =1/\sqrt{1-v^{2}}$ and $w=\rho+p=Ts$ is the
enthalpy density. Equivalently, we have
\begin{equation}
v_{+}v_{-}=\frac{\left( p_{+}-p_{-}\right) }{
\left( \rho _{+}-\rho _{-}\right)  },\
\frac{v_+}{v_{-}}=\frac{ \left( \rho _{-}+p_{+}\right) }{%
 \left( \rho _{+}+p_{-}\right) },
\label{vlandau}
\end{equation}%
which can be readily solved for the velocities in terms of the
thermodynamical quantities.
Notice that these two equations have four unknowns, namely, the velocities $%
v_{\pm }$ and the temperatures $T_{\pm }$ on both sides of the
wall. All other thermodynamical quantities are determined by
the equation of state. In principle, the temperature $T_{+}$ of
the supercooled phase can be calculated using the nucleation
rate. To determine completely the system, one more equation is
needed. This can be obtained from a microscopic calculation of
the friction of the wall with the plasma.

\begin{figure}[htb]
\centering \epsfysize=7cm
\leavevmode \epsfbox{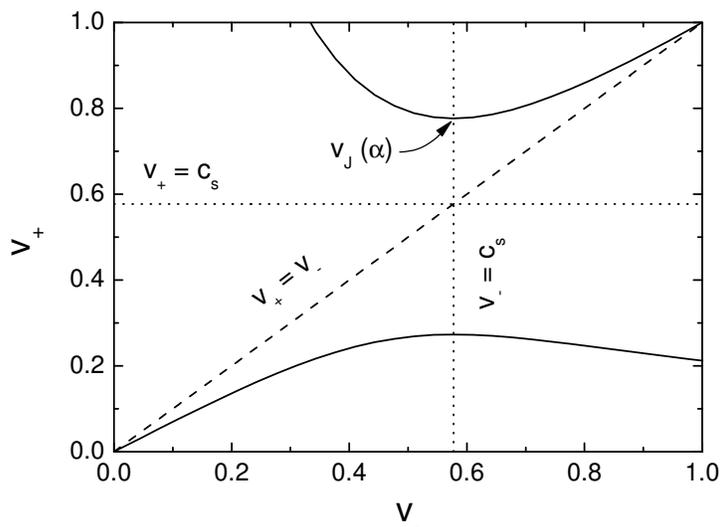}
\caption{$v_{+}$ vs $v_{-}$ according to Eq. (\ref{steinhardt}),
for $\alpha=0.1$}
\label{figvmavme}
\end{figure}

The above equations admit two kinds of solutions, one of them with $
\left\vert v_{-}\right\vert <\left\vert v_{+}\right\vert $, called
detonation, and the other with $\left\vert v_{-}\right\vert
>\left\vert v_{+}\right\vert $, called deflagration (Fig. \ref{detodefla}).
This can be seen by using a simple model in which the
low-temperature EOS is that of radiation, $p_{-}=\rho _{-}/3$,
and the high-temperature EOS is that of radiation plus vacuum
energy, i.e., $\rho _{+}=\tilde{\rho} _{+}+\varepsilon $, with
$p_{+}=\tilde{\rho} _{+}/3-\varepsilon $. From Eqs.
(\ref{eqlandau}) or (\ref{vlandau}) one obtains \cite{s82}
\begin{equation}
v_{+}=\frac{\frac{1}{6v_{-}}+\frac{v_{-}}{2}\pm \sqrt{\left( \frac{1}{6v_{-}}%
+\frac{v_{-}}{2}\right) ^{2}+\alpha ^{2}+\frac{2}{3}\alpha -\frac{1}{3}}}{%
1+\alpha },  \label{steinhardt}
\end{equation}%
where $\alpha =\varepsilon /\tilde{\rho} _{+}$. The solutions
with the plus and the minus signs correspond to detonations and
to deflagrations, respectively. We plot both in Fig.
\ref{figvmavme} for a given value of $\alpha$.  It turns out
that, for a detonation, we have $\left\vert v_{+}\right\vert
>c_{s+}$, and for a deflagration, $\left\vert v_{+}\right\vert
<c_{s+}$, where $c_{s\pm }$ is the speed of sound on each side
of the wall. For the above simple equations of state, the speed
of sound in both phases is that of a relativistic plasma,
$c_{s\pm }=\left( dp_{\pm }/d\rho _{\pm }\right)
^{1/2}=1/\sqrt{3}$.

\begin{figure}[htb]
\centering \epsfxsize=8cm
\leavevmode \epsfbox{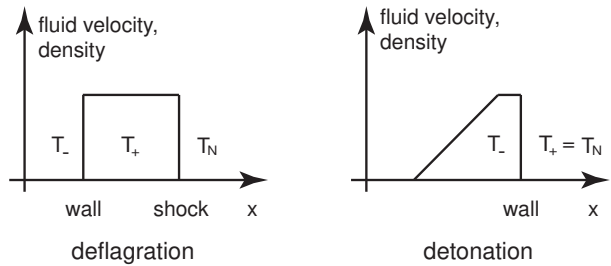}
\caption{Schematically, the profile of the density or the velocity of
the fluid, induced by the moving wall.}
\label{figperfil}
\end{figure}

In the reference frame where the matter far in front of the
bubble wall is at rest and the wall moves, the matter at the
center of the bubble (far behind the wall) must be also at
rest. As a consequence, a single front (the phase transition
front) is not sufficient to satisfy all the boundary
conditions. As a result, it turns out that a deflagration front
must be preceded by a shock front (Fig. \ref{figperfil}, left).
Between the shock and the deflagration fronts, the matter is
compressed and has a finite velocity. Outside this region the
matter is at rest. On the other hand, a detonation front hits
fluid which is at rest, the density increases suddenly, and the
phase transition front is followed by a rarefaction wave (Fig.
\ref{figperfil}, right). Comparing with the reference frame of
the wall, we see that the deflagration wall velocity is given
by $v_{w}=\left\vert
v_{-}\right\vert $, while the detonation wall velocity is given by $%
v_{w}=\left\vert v_{+}\right\vert .$ According to the inequalities
above, a detonation is always supersonic, $v_{w}>c_{s+}.$ A
deflagration may in principle be supersonic or subsonic (it \emph{is}
subsonic with respect to the fluid in front of it, which is moving in
the direction of the wall).

\subsubsection{Strong and weak solutions}

Depending on the velocity $v_{-}$ of the outflow, deflagrations are
divided into weak ($\left\vert v_{-}\right\vert <c_{s-}$), Jouguet
($\left\vert
v_{-}\right\vert =c_{s-}$) and strong ($\left\vert v_{-}\right\vert >c_{s-}$%
) deflagrations. Thus, strong deflagrations are supersonic, and weak
deflagrations are subsonic. Detonations are also classified into strong ($%
\left\vert v_{-}\right\vert <c_{s-}$), Jouguet ($\left\vert
v_{-}\right\vert =c_{s-}$), and weak ($\left\vert v_{-}\right\vert
>c_{s-}$) detonations. Since the incoming flow is supersonic, strong
detonations perturb the fluid more strongly than weak detonations.

The various kinds of solutions have been extensively
investigated \cite{ikkl94,s82,gkkm84,k85,eikr92}. Using
thermodynamical arguments, one can find regions in parameter
space where each solution is allowed. In particular, the
non-negativity of entropy production may rule out the whole
family of detonations or deflagrations \cite{eikr92}. However,
such arguments do not rule out a particular type (i.e., weak,
Jouguet, or strong) of detonation or deflagration \cite{l94}.
In the case of deflagrations, in Ref. \cite{l94} it is argued
that strong deflagrations are mechanically unstable, although
in principle there is no physical reason against their
existence. A special type of supersonic Jouguet deflagration,
which is followed by a rarefaction wave, was shown to exist in
Ref. \cite{kl95}.

In the case of detonations, it is known \cite{landau} that, for
the case of chemical burning, detonation bubbles can only grow
as Jouguet detonations. Remarkably, this condition (which is
referred to as the Chapman-Jouguet hypothesis) means that a
microscopic calculation of the wall velocity  is irrelevant,
since the system is completely determined by the
energy-momentum conservation and the boundary conditions. In
Ref. \cite{s82} it was argued that the Chapman-Jouguet
hypothesis should be valid also for cosmological phase
transitions. Setting $v_-=c_s$ in Eq. (\ref{steinhardt}) leads
to a very simple formula $v_w=v_J(\alpha)$ for the wall
velocity. However, it was shown in Ref. \cite{l94} that, in
contrast to the case of chemical burning, the Chapman-Jouguet
hypothesis does not hold in the case of phase transitions. It
was also shown in Refs. \cite{s82,l94} that strong detonations
are impossible, since they cannot satisfy the boundary
condition of a vanishing fluid velocity at some point behind
the bubble wall. Nevertheless, weak detonations are possible,
and the velocity will be determined in general by microscopical
processes. Thus, in a realistic model the detonation velocity
will depend on a friction parameter $\eta $ as well as on the
hydrodynamic parameter $\alpha $. The wall velocity will always
satisfy the condition $v_{w}\left( \alpha ,\eta \right) \geq
v_{J}\left( \alpha \right) $ (as can be seen in Fig.
\ref{figvmavme}). Hence, $v_{w}$ may give the Jouguet result
only in some limiting cases.

\subsubsection{Detonations vs deflagrations}

It is well known that, if the supercooling is not considerable
and the friction is large enough, the bubbles will grow as
deflagrations, whereas detonations demand  small values of the
friction and more extreme conditions (i.e., strong
supercooling) \cite{ikkl94,gkkm84,eikr92}. It is often argued,
though, that the propagation of the deflagration wall becomes
turbulent and eventually turns into a detonation due to shape
instabilities of the wall. As discussed in Ref. \cite{l92} for
the case of the QCD phase transition and in Ref. \cite{kf92}
for the case of the electroweak phase transition, small
wavelength perturbations of the deflagration front are
stabilized by the interface tension, while large ones grow
exponentially. However, a more detailed study \cite{hkllm93}
shows that the strong dependence of the wall velocity on
temperature actually stabilizes the large scale perturbations.
The stability of the deflagration front thus depends on several
quantities (these are essentially the latent heat, the value of
the wall velocity and its dependence on temperature). It was
found, in particular, that the deflagration front is stable
under perturbations \textit{at all scales} for velocities above
a certain critical value $v_{c}$. For instance, in the
(unrealistic) case of the minimal Standard Model with a Higgs
mass $ m_{H}=40GeV$, the critical velocity is $v_{c}\approx
0.07$ \cite{hkllm93}. Below this velocity, perturbations are
unstable on scales larger than a critical scale $\lambda _{c}$.
However, depending on the parameters, the value of $\lambda
_{c}$ may be much larger than the size of bubbles. Furthermore,
the characteristic time for the growth of the instabilities may
be larger than the duration of the phase transition. Therefore,
in the general case, the deflagration front will be probably
stable. A study of the hydrodynamical stability of bubble
growth is beyond the scope of this paper.

\subsection{Microphysics}

In any case, it is clear that the determination of the wall
velocity requires additional input. To consider the effect of
friction together with the fluid equations (\ref{eqlandau}), it
is customary to introduce a damping term of the typical form
$u^{\mu }\partial _{\mu }\phi $ in the equation of motion for
the Higgs field \cite{ikkl94},
\begin{equation}
\partial _{\mu }\partial ^{\mu }\phi +\frac{\partial \mathcal{F}\left( \phi
,T\right) }{\partial \phi }+\left( T_{c}\tilde{\eta}\right) u^{\mu }\partial
_{\mu }\phi =0.
\end{equation}%
Here, the dimensionless damping coefficient $\tilde{\eta}$ is a free
parameter. Nevertheless, we can relate $\tilde{\eta}$ to the actual friction
coefficient obtained from microscopical calculations in a specific model, as
we shall see in section \ref{paramfric}. For simplicity, we shall consider a
planar wall moving along the $z$ axis. In the frame of the wall we have
\begin{equation}
-\phi ^{\prime \prime }+\partial \mathcal{F}/\partial \phi +T_{c}\tilde{\eta}%
u\phi ^{\prime }=0,  \label{eqfieta}
\end{equation}%
where $u=\gamma v$, with $v$ the fluid velocity in the $z$ direction.

Multiplying by $\phi ^{\prime }\left( z\right) $ and integrating over
$-\infty <z<+\infty $, the first term disappears. Then,
\emph{if we neglect the variation of temperature with $z$}, the fluid velocity is $%
v=-v_{w}$, and we obtain
\begin{equation}
\mathcal{F}\left( 0,T\right) -\mathcal{F}\left( \phi _{m},T\right) =T_{c}%
\tilde{\eta}\sigma \gamma v_{w}, \label{eqvwold}
\end{equation}%
where
\begin{equation}
\sigma =\int \left[ \phi ^{\prime }\left( z\right) \right] ^{2}dz
\end{equation}%
is the surface tension of the bubble wall.  Equation (\ref{eqvwold})
gives the well known nonrelativistic approximation for the wall
velocity
\begin{equation}
v_{w}\approx \Delta \mathcal{F}\left( T\right) /\eta ,  \label{vwmicro}
\end{equation}%
where
\begin{equation}
\eta =\tilde{\eta}T_{c}\sigma ,  \label{eta}
\end{equation}%
and  $\Delta \mathcal{F}$ is the free energy density difference
$\mathcal{F}_{+}\left(
T\right) -\mathcal{F}_{-}\left( T\right) $, which gives the pressure difference $%
\Delta p=p_{-}\left( T\right) -p_{+}\left( T\right) $ between the two phases.

If we now take into account hydrodynamics, we can still integrate Eq. (\ref%
{eqfieta}), but we cannot ignore the fact that the temperature and velocity
of the fluid depend on $z$. Before multiplying by $\phi ^{\prime }$ and
integrating, we use the equation $\partial \mathcal{F}/\partial \phi =d%
\mathcal{F}/d\phi -\left( \partial \mathcal{F}/\partial T\right) \left(
\partial T/\partial \phi \right) $ \cite{ikkl94}. Then, proceeding as
before, we obtain
\begin{equation}
\mathcal{F}\left( 0,T_{+}\right) -\mathcal{F}\left( \phi _{m}\left(
T_{-}\right) ,T_{-}\right) -\int_{T_{-}}^{T_{+}}\frac{\partial \mathcal{F}}{%
\partial T}dT+\tilde{\eta}T_{c}\int_{-\infty }^{+\infty }\left[ \phi
^{\prime }\left( z\right) \right] ^{2}u\left( z\right) dz=0.
\end{equation}%
The first two terms are just $\mathcal{F}_{+}\left( T_{+}\right) -\mathcal{F}%
_{-}\left( T_{-}\right) =p_{-}-p_{+}$. To simplify the calculations, some
approximations are necessary. In the first integral, the function $\partial
\mathcal{F}/\partial T\left( \phi \left( x\right) ,T\left( x\right) \right) $
varies from
\begin{equation}
\frac{\partial \mathcal{F}}{\partial T}\left( \phi _{m}\left( T_{-}\right)
,T_{-}\right) =\frac{d\mathcal{F}_{-}}{dT}\left( T_{-}\right) =-s_{-}\left(
T_{-}\right)
\end{equation}%
to
\begin{equation}
\frac{\partial \mathcal{F}}{\partial T}\left( 0,T_{+}\right) =\frac{d\mathcal{F}_{+}%
}{dT}\left( T_{+}\right) =-s_{+}\left( T_{+}\right)
\end{equation}%
(we have used the fact that $\phi_m$ and 0 are minima of
$\mathcal{F}$). If $\phi \left( x\right) $ and $T\left( x\right) $
change smoothly, we expect that a linear function will give a good
approximation to the integrand $\partial \mathcal{F}/\partial T$
inside the thin wall. Therefore, the integral yields $-\left(
s_{-}+s_{+}\right) \left( T_{+}-T_{-}\right) /2$. For the second
integral, it is a good approximation to assume that $\phi ^{\prime
2}$ is symmetric around the center of the wall at $z=0$; hence, this
can be approximated by $\sigma \left( u_{+}+u_{-}\right) /2$
\cite{ikkl94}. Hence, we have
\begin{equation}
p_{+}-p_{-}-\frac{1}{2}\left( s_{+}+s_{-}\right) \left( T_{+}-T_{-}\right) +%
\frac{\eta }{2}\left( v_{+}\gamma _{+}+v_{-}\gamma _{-}\right) =0,
\label{eqmicro}
\end{equation}%
where $\eta =\tilde{\eta}T_{c}\sigma $ as before.

Notice that the approximations we have made to obtain Eq.
(\ref{eqmicro}) do not involve any assumption on the wall velocity
nor on the equation of state. Hence, this equation can be used for
the treatment of relativistic as well as nonrelativistic velocities,
and is model-independent. Provided the various thermodynamical
quantities are related by means of an equation of state, Eqs.
(\ref{eqlandau}) and (\ref{eqmicro}) can be solved to obtain the
velocities $v_{-}$ and $v_{+}$ as functions of the temperature
$T_{+}$ in front of the wall. The result will depend on parameters of
the theory that appear in the EOS, such as the latent heat, and on
the friction coefficient. In practice, however, it is very difficult
to solve these equations unless we use a simple model.

\subsection{The equation of state}

As we have seen, the relation (\ref{steinhardt}) between $v_{+}$ and $v_{-}$
can be derived from Eqs. (\ref{eqlandau}), using the relations
\begin{equation}
p_{-}=\rho _{-}/3  \label{prome}
\end{equation}%
for the low-T phase, and
\begin{equation}
\rho _{+}=\tilde{\rho} _{+}+\varepsilon ,\ p_{+}=\tilde{\rho} _{+}
/3-\varepsilon  \label{proma}
\end{equation}%
for the high-T phase. In order to obtain a second relation
between $v_+$ and $v_-$ from Eq. (\ref{eqmicro}), the EOS for
each phase has to be more specific. The \emph{thermal} energy
densities $\rho _{-}\left( T\right) $ and $\tilde{\rho}
_{+}\left( T\right) $ must be of the form $aT^{4}$, where
$a=\pi^2g/30$ and $g$ is the number of relativistic degrees of
freedom. Hence, the simplest possibility is to assume that both
phases have the same $a$. Such a simple model was considered in
Ref. \cite{s82}. This model, however, fails to describe a
realistic phase transition if the vacuum energy $\varepsilon $
is a constant. Indeed, the critical temperature is defined as
that $T$ at which the pressures of the two phases are equal,
$p_{+}\left( T_{c}\right) -p_{-}\left( T_{c}\right) =0$.
However, with $p_{+}=aT^{4}/3-\varepsilon $ and
$p_{-}=aT^{4}/3,$ we have $p_{+}\left( T\right) -p_{-}\left(
T\right) =-\varepsilon $ at any $T$. This would be possible if
$\varepsilon $ were a function of $T$. With $\varepsilon$
constant, this model does not have a critical temperature.

The existence of a critical temperature $T_c$, below which
bubbles nucleate and grow, constitutes an important distinction
between the case of a phase transition and that of chemical
burning. In a phase transition, supercooling is needed for
bubbles to grow. In the first place, the temperature must
\emph{decrease sufficiently} below $T_c$ so that bubbles
nucleate. In the second place, even if for some reason (e.g.,
by means of inhomogeneous nucleation) bubbles nucleate at
$T=T_c$, the condition $T<T_c$ is still necessary to achieve a
pressure difference $p_-(T)>p_+(T)$, so that bubbles can grow.
In the case of chemical burning, the speed of the reaction
increases with temperature, and the temperature must \emph{rise
sufficiently} for the combustion to proceed. If the reaction is
strongly exothermic, it is sufficient to rise the temperature
at a single point. The heat that is released by the reaction at
that point rises the temperature of the surrounding gas, and
the reaction may extend over the whole gas \cite{landau}. As a
consequence, the higher the released energy, the larger the
velocity of the combustion front. In the case of a phase
transition, the release of latent heat is not necessary for the
process to continue. On the contrary, the reheating of the
supercooled gas \emph{slows down} the phase transition, as the
temperature approaches $T_c$.\footnote{The process we have just
described corresponds to a deflagration or slow combustion. For
the detonation case one can argue similarly. Essentially, in
the case of chemical burning, the combustion takes place behind
the detonation front because the temperature is risen by the
front. Evidently, a phase transition does not need such a
temperature rise.}

To keep the EOS as simple as possible, we shall consider $\varepsilon
$ constant, and different values $a_{+}$ and $a_{-}$  in the two
phases. This gives the well known {bag equation of state}, which is
appropriate to a system with a first-order phase transition:
\begin{eqnarray}
\rho _{+} =a_{+}T^{4}+\varepsilon , & &p_{+}=a_{+}T^{4}/3-\varepsilon ,
\label{eos} \\
\rho _{-} =a_{-}T^{4}, & &p_{-}=a_{-}T^{4}/3,  \nonumber
\end{eqnarray}
and the entropy is given by $s_{\pm }=\frac{4}{3}a_{\pm
}T^{3}$. Since Eq. (\ref{steinhardt}) was derived using the
relations (\ref{prome},\ref{proma}), the result is still valid
as a function of $\alpha =\varepsilon /\tilde{\rho} _{+}$, with
$\tilde{\rho} _{+}=a_{+}T_{+}^{4}$. However, the parameter
$\varepsilon $ has a different meaning in each model. Indeed,
with $a_{+}=a_{-}$ the energy released in the phase transition
is just $\varepsilon $. In a realistic model the vacuum energy
$\varepsilon $ does not coincide with the latent heat $L=\rho
_{+}\left( T_{c}\right) -\rho _{-}\left( T_{c}\right) $, since
entropy is released together with vacuum energy. For the bag
equation of state, the critical temperature is readily obtained
by equating the pressures $p_{+}$ and $p_{-}$. We have
\begin{equation}
\left( a_{+}-a_{-}\right) T_{c}^{4}=3\varepsilon ,  \label{tc}
\end{equation}%
from which we obtain the latent heat%
\begin{equation}
L=4\varepsilon .  \label{leps}
\end{equation}%
In fact, in a realistic model, the vacuum energy density does
not vanish immediately after the phase transition. Hence, we
will have in general two parameters $ \varepsilon _{\pm }$, one
for each phase. Nevertheless, this only amounts to replacing
$\varepsilon =\varepsilon _{+}-\varepsilon _{-}$ in Eqs.
(\ref{tc},\ref{leps}) and in the definition of $\alpha $. The
expression of $\alpha $ in terms of the latent heat, $\alpha
=L/(4\tilde{\rho} _{+}),$ remains unchanged. For simplicity, we
shall set $ \varepsilon _{-}=0$ in the following.

Notice that the positivity of the pressure at $T=T_{c}$ (i.e.,
$p_{+}=p_{-}>0$) implies that the ratio
\begin{equation}
\alpha _{c}\equiv \varepsilon /\left( a_{+}T_{c}^{4}\right)  \label{ac}
\end{equation}
cannot be arbitrarily large, namely, $\alpha _{c}<1/3$. This relation
between the vacuum and the thermal energy densities is valid beyond this
simple model \cite{ms08}.

Using  the fluid conditions (\ref{eqlandau}) and the equations
of state (\ref{eos}), the friction equation (\ref{eqmicro})
becomes
\begin{equation}
\frac{p_{+}-p_{-}}{\tilde{\rho} _{+}}-\frac{2}{3}\left( 1+\frac{s_{-}}{
s_{+}}\right) \left( 1-\frac{T_{-}}{T_{+}}\right) +\frac{2\alpha _{+}\eta }{L%
}\left( \left\vert v_{+}\right\vert \gamma _{+}+\left\vert v_{-}\right\vert
\gamma _{-}\right) =0,  \label{eqfric}
\end{equation}%
with
\begin{equation}
\frac{p_{+}-p_{-}}{\tilde{\rho} _{+}}=\alpha _{+}\frac{4v_{+}v_{-}}{%
1-3v_{+}v_{-}},  \label{deltap}
\end{equation}%
\begin{equation}
\frac{s_{-}}{s_{+}}=\frac{a_{-}}{a_{+}}\left( \frac{T_{-}}{T_{+}}\right)
^{3},\quad \frac{T_{-}}{T_{+}}=\left( \frac{a_{+}}{a_{-}}\frac{\rho _{-}}{%
\tilde{\rho} _{+}}\right) ^{1/4},  \label{st}
\end{equation}%
and
\begin{equation}
\frac{\rho _{-}}{\tilde{\rho} _{+}}=1-\alpha _{+}\frac{1+v_{+}v_{-}}{%
1/3-v_{+}v_{-}},  \label{romeroma}
\end{equation}%
where we have used instead of $\alpha$ the notation
\begin{equation}
\alpha _{+}\equiv \varepsilon /(a_{+}T_{+}^{4}).
\end{equation}
The velocities are further related by Eq. (\ref{steinhardt}),
with the + sign for detonations and the - sign for
deflagrations, and the ratio $a_{-}/a_{+}$ is a parameter of
the model. Using Eqs. (\ref{tc}) and (\ref{ac}) we can write
\begin{equation}
a_{-}/a_{+}=1-3\alpha _{c}.  \label{ameama}
\end{equation}%
It can be seen from Eqs. (\ref{vlandau}) that $v_{+}v_{-}\leq
1/3$. Hence, according to Eq. (\ref{deltap}), we have
$p_{+}\left( T_{+}\right) -p_{-}\left( T_{-}\right) >0$. This
means that the temperature difference
that is established around the wall inverts \emph{locally} the relation $%
p_{-}\left( T\right) >p_{+}\left( T\right) ,$ which holds for $T<T_{c}$.

We can solve the above equations for $v_{+}$ or $v_{-}$ as a
function
of $T_{+}$ (or, equivalently, $\alpha _{+}$) and the parameters $\eta $ and $%
\alpha _{c}$. In the case of detonations, the temperature
$T_{+}$ in front of the bubble wall is just the temperature
$T_{N}$ at which bubbles nucleate, so we have
\begin{equation}
\alpha _{+}=\alpha _{N}\equiv \frac{\varepsilon }{a_{+}T_{N}^{4}}\qquad
\mathrm{(detonations)},  \label{amndeto}
\end{equation}
and the fluid on that side of the wall is at rest, so
\begin{equation}
v_{w}=|v_{+}|.  \label{vmwdeto}
\end{equation}
Thus, we only need to solve Eqs. (\ref{eqfric}-\ref{romeroma}) to
obtain $v_{w}$. This makes this case in general simpler than the
deflagration case\footnote{%
If we wanted to find the temperature or velocity profiles inside the
bubble, then we would have to consider the equations for the fluid
behind the wall and impose boundary conditions at the center of the
bubble.}.

In the case of deflagrations, we must relate the temperature
$T_{+}$ of the fluid in the  reheated region between the
phase-transition front and the shock front, to the nucleation
temperature $T_{N}$ of the fluid beyond the shock front (see
Fig. \ref{figperfil}). To do that, we must consider the fluid
conditions (\ref{eqlandau}) for the shock front discontinuity.
These are simpler than those for the wall, since we have the
same phase on both sides of this front. Calling $v_{1}$ the
velocity of the fluid in front of the shock and $v_{2}$ the
velocity behind it, we have, in the reference frame of the
shock front \cite{landau},
\begin{equation}
\left\vert v_{1}\right\vert =\frac{1}{\sqrt{3}}\left( \frac{%
3T_{+}^{4}+T_{N}^{4}}{3T_{N}^{4}+T_{+}^{4}}\right) ^{1/2},\ \left\vert
v_{2}\right\vert =\frac{1}{3\left\vert v_{1}\right\vert }.
\end{equation}%
In the ``laboratory'' frame, the fluid is at rest behind the phase
transition discontinuity, and also ahead of the shock front, so we
have
\begin{equation}
v_{w}=\left\vert v_{-}\right\vert ,  \label{vmwdefla}
\end{equation}
whereas $v_{1}$ gives the velocity of the shock front, $v_{1}=-v_{\mathrm{sh}%
}$. The velocity of the fluid $v_{\mathrm{fl}}$ between the two fronts can
be computed from the velocities $v_{+}$ and $v_{-}$, or from $v_{1}$ and $%
v_{2}$. Equating both results we find the relation between $T_{+}$ and $T_{N}
$,
\begin{equation}
\frac{\sqrt{3}\left( \alpha _{N}-\alpha
_{+}\right) }{\sqrt{\left( 3\alpha _{N}+\alpha _{+}\right) \left( 3\alpha
_{+}+\alpha _{N}\right) }}=\frac{v_{+}-v_{-}}{1-v_{+}v_{-}} \qquad
\mathrm{(deflagrations)}.  \label{amndefla}
\end{equation}

Thus, we can find $v_{w}\left( \alpha _{c},\alpha _{N},\eta
\right) $, both in the case of a detonation and of a
deflagration. Notice that the parameter $\alpha _{c}$ gives the
ratio of latent heat to thermal energy density at $T=T_{c}$.
Thus, for $\alpha _{c}$ fixed, the parameter $\alpha _{N}$
gives a measure of the amount of supercooling, since $\alpha
_{N}/\alpha _{c}=\left( T_{c}/T_{N}\right) ^{4}$. We have seen
that $0<\alpha _{c}<1/3$. On the other hand, $\alpha _{N}$ is
only bounded below, $\alpha _{N}>\alpha _{c}$. Nevertheless, in
general we will have $T_{N}\sim T_{c}$ and $\alpha _{N}\lesssim
1$. We could have $\alpha _{N}\gg \alpha _{c}$ only in a phase
transition with extremely strong supercooling.

\section{Numerical results} \label{nume}

We have solved numerically Eqs. (\ref{eqfric}-\ref{romeroma})
and (\ref{steinhardt}), with Eqs. (\ref{amndeto},\ref{vmwdeto})
for detonations and Eqs. (\ref{vmwdefla},\ref{amndefla}) in the
case of deflagrations. In this section we present the numerical
results, considering independent variations of the parameters
$\alpha _{c}$, $\alpha _{N}$ and $\eta $. Nevertheless, we
shall make variations around two reference sets of parameters,
corresponding to different points in the physical model of
section \ref{model} (indicated by crosses in Fig.
\ref{figmodel}). We choose $\alpha _{c}\approx 1.27\times
10^{-2}$, $\alpha _{N}\approx 1.48\times 10^{-2}$, $\eta
/L\approx 4.03\times 10^{-2}$, which are obtained when the
fundamental parameters of the model are such that the phase
transition is rather strongly first-order, i.e., the order
parameter is $\phi _{N}/T_{N}\approx 2$; and $\alpha
_{c}\approx 3.19\times 10^{-3}$, $\alpha _{N}\approx 3.25\times
10^{-3}$, $\eta /L\approx 2.48\times 10^{-2}$, obtained from a
weaker phase transition, but still with $\phi _{N}/T_{N}\approx
1$.

\subsection{Detonations}

\begin{figure}[htb]
\centering \epsfysize=6cm
\leavevmode \epsfbox{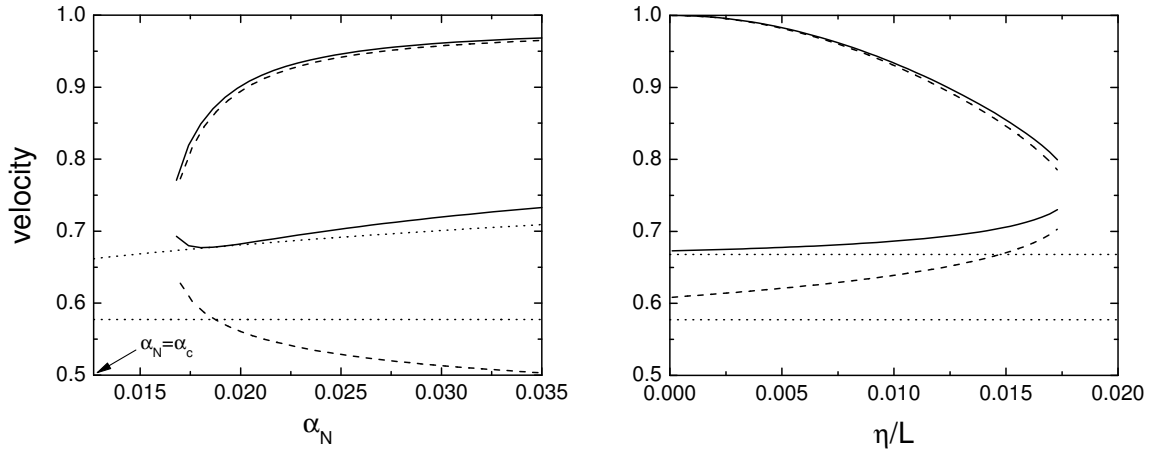}
\caption{Detonation solutions for $\alpha _{c}\approx 1.27\times
10^{-2}$. Left: the velocities as functions of
$\alpha_N$ for $\eta /L\approx
4.03\times 10^{-2}$. Right: the velocities as functions of
$\eta/L$ for $\alpha _{N}\approx 1.48\times 10^{-2}$.
Solid lines indicate the velocity $v_w$ of the
wall, dashed lines the velocity $v_-$ of the outgoing fluid, and dotted lines
the speed of sound (lower curves) and the Jouguet velocity (upper curves).}
\label{figdeto}
\end{figure}

Let us consider a strong phase transition ($\phi _{N}/T_{N}\approx
2$). In Fig. \ref{figdeto} (left panel), we have fixed the values of
$\alpha _{c}$ and $\eta /L$, and we have varied $\alpha _{N}$,
starting from $\alpha _{N}=\alpha_c$, i.e., we have decreased the
supercooling temperature $T_{N}$ below $T_{c}$.
For $\alpha _{N}$ close to $\alpha _{c}$, no detonation
solutions exist (i.e., some supercooling is required). When
$\alpha _{N}$ is large enough we find two solutions for the
detonation wall velocity (solid lines). One of them (lower
solid curve) is close to the Jouguet velocity $v_{J}\left(
\alpha _{N}\right)$, which is indicated by the upper dotted
line. This solution, however, does not seem to be physical. In
the first place, the velocity \emph{decreases with the
supercooling} until it reaches the Jouguet point. Beyond this
point, the velocity increases, but the solution has become a
strong detonation, since the velocity $v_{-}$ (lower dashed
curve) crosses below the speed of sound (indicated by the lower
dotted line). The other solution, in contrast, is always a weak
detonation (the corresponding velocity $v_-$ is indicated by
the upper dashed curve). As a matter of fact, this detonation
is quite weak, as the outflow velocity $v_-$ is very close to
the inflow velocity $v_+=v_w$, which implies that the fluid is
almost unperturbed by the wall.

This behavior can also be seen if we fix $\alpha _{c}$ and
$\alpha _{N}$, and plot the wall velocity as a function of
$\eta $ (Fig. \ref{figdeto}, right panel). If the friction is
strong enough, no detonation solution exists. For lower values
of the friction parameter, we have two solutions, but the lower
branch corresponds to a velocity which decreases as the
friction is decreased. Depending on the values of the
parameters, this solution eventually becomes a strong
detonation for low values of $\eta $. In any case, as can be
seen in the figure, the lower branch solution is much stronger
than that of the upper branch. For these reasons, we shall
discard this solution.

\subsection{Deflagrations}

\begin{figure}[htb]
\centering \epsfysize=6cm
\leavevmode \epsfbox{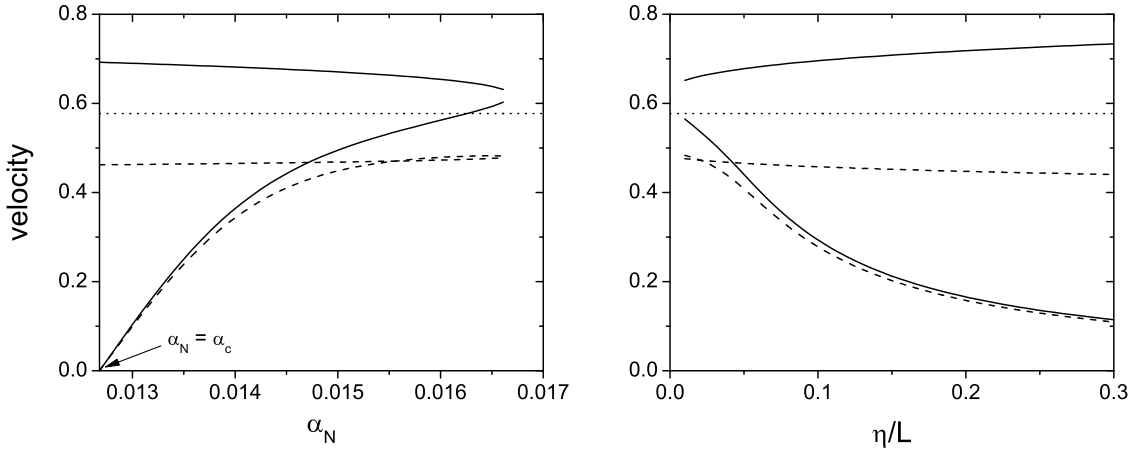}
\caption{Deflagration solutions for the same case as Fig. \ref{figdeto}.
Solid lines indicate the velocity $v_w$ of the
wall, dashed lines the velocity $v_+$ of the incoming fluid, and dotted lines
the speed of sound.}
\label{figdefla}
\end{figure}

Solving numerically Eqs. (\ref{eqfric}-\ref{romeroma}),
(\ref{steinhardt}), and (\ref{amndefla}), we find again two
solutions. However, one of them must be discarded, since it has
a wrong behavior as a function of the parameters: the velocity
decreases with the supercooling and increases with the
friction, and is in general supersonic. The correct solution,
on the other hand, goes to zero for small supercooling and
strong friction. As an illustration, we plot both solutions in
Fig. \ref{figdefla} for the case of a strong phase transition,
as functions of $\alpha_N$ and $\eta $. In contrast to the
detonation case, the deflagration solution always exists in the
limit of small supercooling, $\alpha_N\to\alpha_c$. On the
other hand, the deflagration solution may not exist when the
supercooling becomes too strong or when the friction becomes
too low, as this example illustrates.

\begin{figure}[htb]
\centering \epsfysize=6cm
\leavevmode \epsfbox{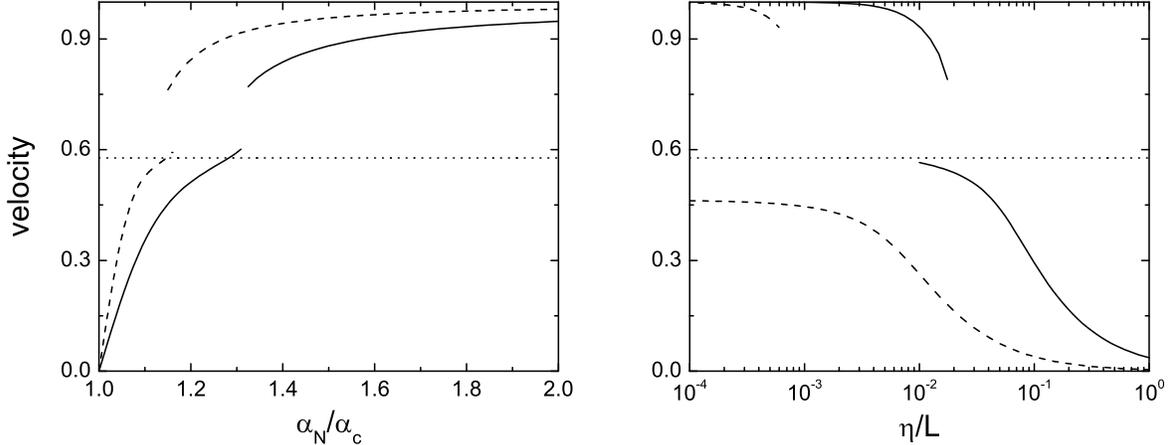}
\caption{Deflagration and detonation solutions as functions of
$\alpha_N$ (left)
and of $\eta/L$ (right) for two families of phase transitions.}
\label{figdetodefla}
\end{figure}

In the high supercooling and low friction range, we have the
detonation solution. We have plotted the physical detonations
and deflagrations together in Fig. \ref{figdetodefla}. Solid
lines correspond to the case of Figs. \ref{figdeto} and
\ref{figdefla}, and dashed lines correspond to variation of
parameters beginning from a weaker phase transition (with $\phi
_{N}/T_{N}\approx 1$). As can be seen in the right panel, for
$\alpha_c$ and $\alpha_N$ fixed, a stronger phase transition
gives, for the same friction, a larger velocity (solid line).
On the contrary, if we vary $\alpha_N$ (left panel), the solid
line gives lower velocities. This is because in this plot we
are considering the the same amount of supercooling for the two
models. Therefore, the model with a larger value of $\alpha_c$
(i.e., of $L$) gives a lower velocity. Indeed, a larger $L$
causes a larger reheating, thus slowing down the phase
transition (both in the deflagration and in the detonation
case).

\begin{figure}[hbt]
\centering \epsfysize=7cm \leavevmode \epsfbox{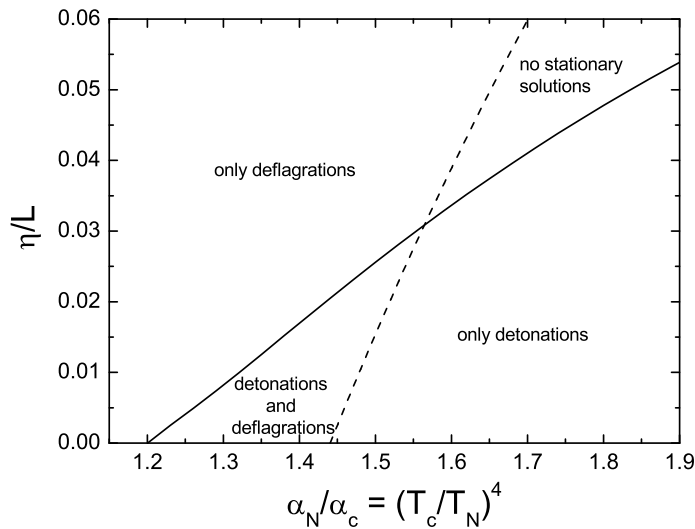}
\caption{The regions of the parameters $\eta$ and $\alpha_N$, for
$\alpha_c=0.05$, where detonations and deflagrations exist. Solid
line: maximum
friction for detonations. Dashed line: minimum friction
for deflagrations.} \label{figregi1}
\end{figure}

\begin{figure}[hbt]
\centering \epsfysize=7cm
\leavevmode \epsfbox{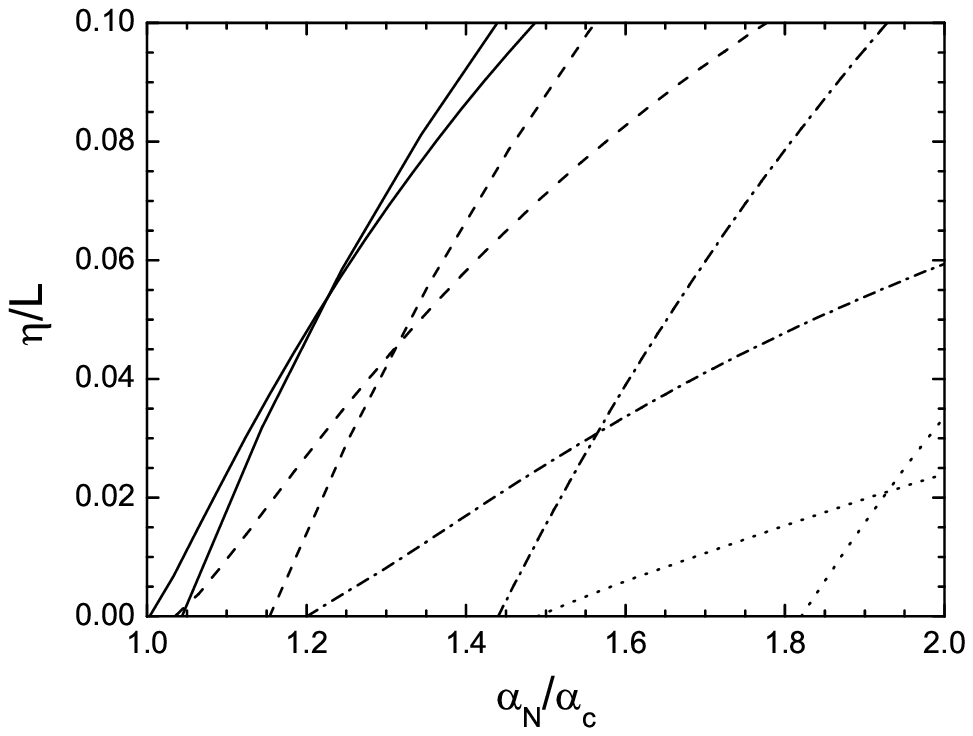}
\caption{The four regions in the plane ($\alpha_N,\eta$)
(as in figure \ref{figregi1}) for
$\alpha_c=10^{-3}$ (solid), $10^{-2}$ (dashed),
$5\times 10^{-2}$ (dashed-dotted),
and $10^{-1}$ (dotted).}
\label{figregi2}
\end{figure}

In Fig. \ref{figdetodefla} we see that different situations may
arise as the parameters are varied. There are ranges in which
detonations and deflagrations coexist, and there are values of
$\alpha_N$ and $\eta $ for which neither solution exists.
Fixing $\alpha_c$, we have calculated the maximum friction for
which detonations exist, and the minimum friction that permits
the existence of deflagrations. We have plotted these limiting
values in Fig. \ref{figregi1} (solid and dashed lines,
respectively). Since the curves cross, the plane is divided
into four regions, in which there exist either deflagrations,
detonations, both, or none. In the latter case, no stationary
state  of bubble expansion is reached. In the case in which
both solutions exist, which of them is realized will depend on
the initial conditions and on the stability of each solution.
Figure \ref{figregi2} shows the four regions for different
values of $\alpha_c$.

It is difficult to determine in general which of these situations may
arise in a physical case. For that aim we would need some relation
between the parameters $\eta/L$, $\alpha_N$, and $\alpha_c$. In the
next section we explore the physical case by considering a particular
model, for which we calculate the relevant parameters $\eta$, $L$,
and $T_N$ as functions of the fundamental parameters of the model.
Nevertheless, some general conclusions can be drawn from Fig.
\ref{figregi1}. Notice that the key parameter for the existence of
detonations is the supercooling parameter $\alpha_N/\alpha_c$: even
for $\eta=0$, detonations exist only if $\alpha_N/\alpha_c$ is large
enough. On the contrary, deflagrations exist up to a certain amount
of supercooling, even for $\eta=0$. Therefore, in a phase transition
with small supercooling, we will have only deflagrations, no matter
how small the friction may be.  This is because hydrodynamics brakes
the propagation of the wall, thus acting effectively as a friction.
This can also be seen in Fig. \ref{figregi2}, by the fact that, as we
increase $\alpha_c$ (i.e., $L$), the deflagration region grows and
the detonation region reduces.

\begin{figure}[htb]
\centering \epsfysize=7cm
\leavevmode \epsfbox{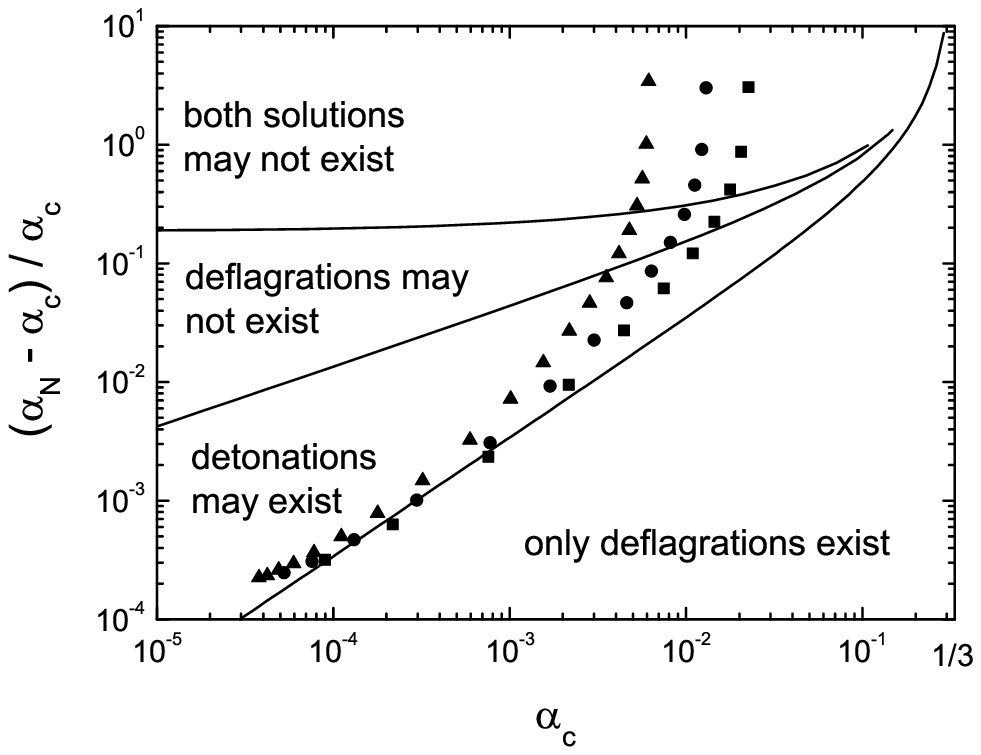}
\caption{The regions of the plane ($\alpha_c,\alpha_N$) where detonations
and deflagrations may or may not exist.
The points correspond to the model of section \ref{model}, for different
values of the d.o.f. $g$ and coupling $h$.
Triangles correspond to $g=2$, circles to $g=6$, and squares to $g=12$.}
\label{figexist}
\end{figure}

Observing Figs. \ref{figregi1} or \ref{figregi2} we see that,
setting $\eta=0$ we find, for each $\alpha_c$, the minimum
$\alpha_N$ for which detonations \emph{may} exist and the
minimum value for which deflagrations \emph{may} not exist.
Besides, the value at which the detonation and deflagration
curves cross defines the minimum $\alpha_N$ for which there
could be no solution. Fig. \ref{figexist} shows these three
values of $\alpha_N$ as functions of $\alpha_c$. These curves
divide the ($\alpha_c,\alpha_N$) plane in four regions. In the
lower region, detonations do not exist, independently of the
value of $\eta$; in the next region detonations will exist if
the value of $\eta$ is small enough; in the third region,
deflagrations will not exist if $\eta$ is too small; in the
upper region, neither solution will exist in a certain range of
the friction parameter. Deflagrations always exist in the two
lower regions and, for large enough friction, in all the
($\alpha_c,\alpha_N$) plane.

\section{The wall velocity in a physical model \label{model}}

So far we have calculated the wall velocity as a function of the
latent heat $L$, the supercooling temperature $T_{N}$, and the
friction coefficient $ \eta $. In a physical model these quantities
are not independent. However, the simple EOS (\ref{eos}) does not
allow to calculate $T_{N}$ nor $\eta $. We shall now consider a model
which permits to obtain these quantities as functions of a single
fundamental parameter. The model we seek should be as simple as
possible, and allow for phase transitions of different strength as
the parameters are varied. It is well known that the electroweak
phase transition is only a smooth crossover in the minimal Standard
Model (SM), whereas many extensions of the model give a strongly
first-order phase transition. Therefore, we choose the simplest
extension of the SM, namely, adding to the model a gauge singlet
scalar $S$ with zero vacuum expectation value
\cite{dhss91,dhs92,ah92}. Besides the coupling to the Higgs, which is
of the form $2h^{2}S^{\dag }SH^{\dag }H$, the field $S$ may have a
mass term $\mu ^{2}S^{\dag }S$ and a quartic term $\lambda _{S}\left(
S^{\dag }S\right) ^{2} $. We will ignore the possibility that cubic
terms exist in the tree-level potential, which may produce a barrier
between minima at $T=0$. Recently, an extension of the SM with
several real singlets $S_{i}$ has been considered \cite{eq07}. These
bosons constitute a hidden sector which couples only to the SM Higgs
doublet through a term $h^{2}H^{\dag }H\sum S_{i}^{2}$. If the fields
$S_{i}$ do not have mass terms, so that they only get a mass from
electroweak breaking, the phase transition can be made exceedingly
strong.

\subsection{The effective potential}

Our model will consist of a scalar field $\phi $ (the background
Higgs field, $\langle H^{0}\rangle \equiv \phi /\sqrt{2}$) with
tree-level potential
\begin{equation}
V_{0}\left( \phi \right) =-m^{2}\phi ^{2}+\frac{\lambda }{4}\phi ^{4},
\label{v0}
\end{equation}%
for which the vacuum expectation value of the Higgs is given by $v=\sqrt{%
2/\lambda }m=246GeV$, and $\lambda $ fixes the Higgs mass, $%
m_{H}^{2}=2\lambda v^{2}$. Imposing the renormalization conditions that the
minimum of the potential and the mass of $\phi $ do not change with respect
to their tree-level values \cite{ah92}, the one-loop zero-temperature
correction is given by
\begin{equation}
V_{1}\left( \phi \right) =\sum_{i}\frac{\pm g_{i}}{64\pi ^{2}}\,\left[
m_{i}^{4}(\phi )\left( \log \left( \frac{m_{i}^{2}(\phi )}{m_{i}^{2}(v)}%
\right) -\frac{3}{2}\right) +2m_{i}^{2}(\phi )m_{i}^{2}(v)\right],  \label{v1loop}
\end{equation}%
where $g_{i}$ is the number of d.o.f. of each particle species,
$m_{i}\left( \phi \right) $ is the $\phi $-dependent mass, and the
upper and lower signs correspond to bosons and fermions,
respectively. The one loop finite-temperature corrections to the free
energy density, including the resummed daisy diagrams are
\begin{eqnarray}
\mathcal{F}_{1}(\phi ,T) &=&\sum_{i}\frac{\pm g_{i}T^{4}}{2\pi ^{2}}
\int_{0}^{\infty }dx\,x^{2}\log \left[ 1\mp \exp \left( -\sqrt{%
x^{2}+m_{i}^{2}\left( \phi \right) /T^{2}}\right) \right] \nonumber \\
&&+\sum_{bosons}\frac{g_{i}T}{12\pi }\left[ m_{i}^{3}\left( \phi \right) -%
\mathcal{M}_{i}^{3}\left( \phi \right) \right] ,   \label{f1loop}
\end{eqnarray}%
where the upper sign stands for bosons, the lower sign stands for fermions,
and $\mathcal{M}_{i}^{2}\left( \phi \right) =m_{i}^{2}\left( \phi \right)
+\Pi _{i}\left( T\right) $, where $\Pi _{i}\left( T\right) $ are the thermal
masses. The last term receives contributions from all the bosonic species
except the transverse polarizations of the gauge bosons. Hence, the one-loop
finite-temperature effective potential is of the form
\begin{equation}
\mathcal{F}(\phi ,T)=V_{0}\left( \phi \right) +V_{1}\left( \phi \right) +
\mathcal{F}_{1}(\phi ,T)+\rho_{\Lambda},  \label{ftot}
\end{equation}
where we have added a constant $\rho_{\Lambda}=-V_0(v)-V_1(v)$,
so that at $T=0$ we will not have a vacuum energy density of
order $v^4$ in the true vacuum. Thus, we have $\rho_-(T=0)=0$.
If all the masses $m_i$ vanish in the symmetric phase, the
constant $\rho_{\Lambda}$ gives the false vacuum energy density
$\rho_+(T=0)$.

For the SM, the relevant contributions come from the $Z$ and $W$ bosons and
the top quark. The top quark contributes with $g_{t}=12$ fermionic d.o.f.,
with a mass $m_{t}\left( \phi \right) =h_{t}\phi /\sqrt{2},$where $h_{t}$ is
the Yukawa coupling ($h_{t}/\sqrt{2}\approx 0.7$). For the weak gauge
bosons, it can be seen numerically \cite{ms08} that a good approximation for
the one-loop correction is to consider a single mass of the form $%
m_{b}\left( \phi \right) =h_{b}\phi $, with $h_{b}\approx 0.35$, and $%
g_{b}=6 $ bosonic d.o.f. The rest of the SM d.o.f. have $h_{i}\ll 1$ and
only contribute a $\phi $-independent term $-\pi ^{2}g_{\mathrm{light}%
}T^{4}/90$, with $g_{\mathrm{light}}\approx 90$. Each extra complex
field $S$ gives contributions to the free energy of the form
(\ref{v1loop},\ref{f1loop}), with $g_{S}=2$ d.o.f. and a mass of the
form $m^{2}\left( \phi \right) =h^{2}\phi ^{2}+\mu ^{2}$. The thermal
mass is given by $\Pi =h^{2}T^{2}/3+\lambda _{S}/4$. For simplicity,
in this work we shall set $ \mu =\lambda _{S}=0$. For illustrative
purposes, we shall also fix the Higgs mass to the value
$m_{H}=125GeV$. It is well known that increasing the Higgs mass
weakens the phase transition. Therefore, higher masses give lower
values of $v_{w}.$ The same happens for nonzero values of the boson
mass parameter $\mu $ since, as $\mu $ is increased, the extra bosons
decouple from the thermal system and the phase transition becomes
weaker. In a forthcoming paper \cite{ms09} we will consider several
particular (and physically motivated) extensions of the SM.

\subsection{Critical temperature and latent heat}

We can find the symmetry-breaking minimum $\phi_m(T)$ for the
free energy (\ref{ftot}) by demanding $\partial
\mathcal{F}(\phi,T)/\partial \phi=0$. Then, as we have seen in
section \ref{intro}, we define the functions
$\mathcal{F}_+(T)=\mathcal{F}(0,T)$ and
$\mathcal{F}_-(T)=\mathcal{F}(\phi_m(T),T)$. From these, we
calculate the critical temperature through the condition
$\mathcal{F}_+(T_c)=\mathcal{F}_-(T_c)$, and the latent heat by
\begin{equation}
L= T_c\left(\mathcal{F}'_-(T_c)-\mathcal{F}'_+(T_c)\right). \label{ele}
\end{equation}
Analytic expressions can be obtained by considering the
high-temperature expansion of the thermal integrals appearing
in (\ref{f1loop}). However, this approximation breaks down for
$m_i(\phi)/T\gg 1$. In our model this corresponds to strong
phase transitions, with $h\phi/T\gg 1$. Therefore, we shall
calculate numerically $T_c$ and $L$.

The energy density of the high-temperature phase can be derived
from the free energy density (\ref{ftot}) by the relation
$\rho_+=\mathcal{F}_+(T)-T\mathcal{F}'_+(T)$. The thermal
energy density is given by
$\tilde{\rho}_+=\rho_+-\rho_{\Lambda}$. In general, we have
$\tilde{\rho}_+\approx \pi^2g_*T^4/30 $, where $g_*$ is the
number of relativistic d.o.f. Thus we can readily calculate the
parameter $\alpha_c=L/\tilde{\rho}_+(T_c)$. In order to
calculate the parameter $\alpha_N=L/\tilde{\rho}_+(T_N)$, we
need to compute the temperature $T_N$ at which bubbles
nucleate. The latent heat parameter $L$ in $\alpha_N$ must be
the same as in $\alpha_c$, since for the bag equation of state
the released energy does not depend on temperature. Therefore,
in order to apply the results of the previous sections, we must
use Eq. (\ref{ele}) for the computation of $\alpha_N$, which
gives the correct relation $\alpha_N/\alpha_c=(T_c/T_N)^4$. If
we used instead the energy that is released at $T=T_N$, which
is larger, we would be overestimating the velocity, since for
the bag EOS this would be equivalent to considering a stronger
supercooling.

\subsection{The Nucleation temperature}

In a first-order phase transition, the nucleation of bubbles is governed by
the three-dimensional instanton action
\begin{equation}
S_{3}=4\pi \int_{0}^{\infty }r^{2}dr\left[ \frac{1}{2}\left( \frac{d\phi }{dr%
}\right) ^{2}+V\left( \phi \left( r\right) ,T\right) \right] ,  \label{s3}
\end{equation}%
where $V(\phi,T)\equiv \mathcal{F}(\phi,T)-\mathcal{F}(0,T)$.
The bounce solution of this action, which is obtained by
extremizing $S_{3},$ gives the radial configuration of the
nucleated bubble, assumed to be spherically symmetric. The
action of the bounce coincides with the free energy of a
critical bubble (i.e., a bubble in unstable equilibrium between
expansion and contraction). The bounce solution obeys the
equation
\begin{equation}
\frac{d^{2}\phi }{dr^{2}}+\frac{2}{r}\frac{d\phi }{dr}=V^{\prime }\left(
\phi \right)  \label{eqprofile}
\end{equation}%
with boundary conditions
\begin{equation}
\frac{d\phi }{dr}\left( 0\right) =0,\ \lim_{r\rightarrow \infty }\phi \left(
r\right) =0.
\end{equation}%
The thermal tunneling probability for bubble nucleation per
unit volume and time is \cite{nucl}
\begin{equation}
\Gamma \left( T\right) \simeq A\left( T\right) e^{-S_{3}\left( T\right) /T},
\label{gamma}
\end{equation}%
with $A\left( T\right) =\left[ S_{3}\left( T\right) /2\pi T\right] ^{3/2}$.

The nucleation temperature $T_{N}$ is defined as that at which the
probability of finding a bubble in a causal volume is 1,
\begin{equation}
\int_{t_{c}}^{t_{N}}dt\Gamma \left( T\right) V=1,
\end{equation}%
where $t_c$ is the time at which the Universe reaches the
critical temperature $T_c$ and $t_N$ is the time at which the
first bubbles are nucleated in a causal volume $V$. In the
radiation-dominated era we have $V\sim \left( 2t\right) ^{3}$,
and the time-temperature relation is given by\footnote{This
relation may change due to reheating during the development of
the phase transition, for $t>t_N$ \cite{h95,ms08,m04}.}
\begin{equation}
dT/dt=-HT,
\end{equation}
where $H$ is the expansion rate, $H=\sqrt{8\pi G\rho_+(T)/3}$.
Here, $G$ is Newton's constant. If $\rho_+\approx
\tilde{\rho}_+\approx \pi^2g_*T^4/30$, then the
time-temperature relation is given by the usual expression
$t=\xi M_P/T^2$, where $M_P$ is the Planck mass and
$\xi=\sqrt{45/(16\pi^3g_*)}$.

The nucleation rate $\Gamma (T)$ can be calculated by solving
numerically Eq. (\ref{eqprofile}) for the bubble profile, then
integrating Eq. (\ref{s3}) for the bounce action and, finally,
using the result in Eq. (\ref{gamma}). Analytical
approximations to Eqs. (\ref{s3}-\ref{gamma}) have large errors
due to the exponential dependence of $\Gamma$ and the
sensitivity of $S_3$ to temperature. We solved Eq.
(\ref{eqprofile}) iteratively by the overshoot-undershoot
method\footnote{See Ref. \cite{ms08} for details.}.

\subsection{The friction coefficient}

The effect of microphysics on the propagation of the bubble
wall can be calculated by considering the equation for the
Higgs field in the hot plasma. From energy conservation
considerations, one can derive the equation
\begin{equation}
\partial _{\mu }\partial ^{\mu }\phi +\frac{\partial \mathcal{F}\left( \phi
,T\right) }{\partial \phi }+\sum_{i}g_{i}\frac{\partial m_{i}^{2}}{\partial
\phi }\int \frac{d^{3}p}{\left( 2\pi \right) ^{3}2E_{i}}\delta f_{i}=0,
\label{eqphi}
\end{equation}%
where the sum is over all particle species that couple with $\phi $,
$m_{i}$ are the $\phi $-dependent masses, and $\delta f_{i}$ are the
deviations from the equilibrium distribution functions, induced by
the motion of the wall. The deviations $\delta f_{i}$ have been
calculated either using kinetic theory \cite{dlhll92,lmt92,k92,mp95},
or considering infrared excitations of bosonic fields \cite{mt97},
which undergo overdamped evolution \cite{m00}.

Calculations of the friction are usually carried out
\emph{ignoring hydrodynamics} (i.e., disregarding temperature
and velocity profiles of the fluid). Assuming stationary motion
in the z direction, the first term in Eq. (\ref{eqphi}) becomes
$\left( v_{w}^{2}-1\right) \phi ^{\prime \prime }\left(
z\right) $, and the deviations $\delta f_{i}$ near the wall
depend on $\phi \left( z\right)$ and $\phi ^{\prime }\left(
z\right) $. To the lowest order, $ \delta f_{i}$ is
proportional to the wall velocity $v_{w}$. As a consequence, if
we multiply by $\phi ^{\prime }\left( z\right) $ and integrate
over $-\infty <z<+\infty $, we obtain
\begin{equation}
\mathcal{F}\left( 0,T\right) -\mathcal{F}\left( \phi _{m},T\right) =\eta
v_{w},  \label{v1}
\end{equation}
where $\eta $ is a friction coefficient which depends on the particle
content of the plasma. For particles with a thermal distribution we
have \cite{js01}
\begin{equation}
\eta _{\mathrm{th}}\sim\sum \frac{3g_{i}h_{i}^{4}}{\left( \Gamma
_{i}/10^{-1}T\right) }\left( \frac{\log \chi _{i}}{2\pi ^{2}}\right) ^{2}
\frac{\phi _{m}^{2}\sigma }{T},  \label{etath}
\end{equation}
where $g_{i}$ is the number of degrees of freedom of species $i$,
$h_{i}$ is the coupling to $\phi$, $\Gamma _{i}$ are interaction
rates which are typically $\lesssim 10^{-1}T$, $\chi _{i}=2$ for
fermions and $\chi _{i}=h_{i}^{-1}$ for bosons, while infrared bosons
give a contribution \cite{m00}
\begin{equation}
\eta _{\mathrm{ir}}\sim\sum\frac{g_{b}}{32\pi }\left( \frac{%
m_{D}}{T}\right) ^{2}\log \left( m_{b}\left( \phi _{m}\right) L_{w}\right)
\frac{T^{3}}{L_{w}},  \label{etair}
\end{equation}%
where $g_{b}$ is the number of bosonic degrees of freedom, $L_{w}$ is
the width of the bubble wall, and $m_{D}$ is the Debye mass, given by $%
m_{D}^{2}\sim h_{b}^{2}T^{2}$, where $h_{b}$ is the coupling to
$\phi$.

For bosonic d.o.f.,  expressions (\ref{etath}) and
(\ref{etair}) dominate in different ranges of the couplings
$h_b$, and we shall assume $\eta=\eta _{\mathrm{th}}+\eta
_{\mathrm{ir}}$. Both expressions involve several
approximations and have $\mathcal{O}(1)$ errors, but should be
parametrically correct.

\subsubsection{Parametrization of the friction \label{paramfric}}

To relate the friction coefficient $\eta $ obtained from
microscopical calculations to the one we used as a free parameter in
the previous sections, compare Eqs. (\ref{v1}-\ref{etair}) with Eqs.
(\ref{vwmicro},\ref{eta}) \cite{m04}. We see that the thermal
friction coefficient (\ref{etath}) can be written in the form $\eta
=\tilde{\eta}T_{c}\sigma $, with
\begin{equation}
\tilde{\eta}_{\mathrm{th}}=\sum \frac{3g_{i}h_{i}^{4}}{\left( \Gamma
_{i}/10^{-1}T\right) }\left( \frac{\log \chi _{i}}{2\pi ^{2}}\right)
^{2}\left( \frac{\phi _{m}}{T}\right) ^{2}.  \label{etatth}
\end{equation}%
The surface tension is roughly given by $\sigma \approx \phi _{m}^{2}/L_{w}.$
Furthermore, in general we have $\phi _{m}\sim T$. Therefore, the last
factor in Eq. (\ref{etair}) becomes $T^{3}/L_{w}\sim T\sigma $. Thus, the
infrared boson contribution can also be written in the form (\ref{eta}),
with
\begin{equation}
\tilde{\eta}_{\mathrm{ir}}=\sum_{\mathrm{gauge}}\frac{g_{b}}{32\pi }\left(
\frac{m_{D}}{T}\right) ^{2}\log \left( m_{b}\left( \phi _{m}\right)
L_{w}\right) .  \label{etatir}
\end{equation}%
Thus, for a given model, we can estimate the value of
$\tilde{\eta}$ in Eq. (\ref{eta}) using Eqs.
(\ref{etatth},\ref{etatir}).

\subsection{Numerical results}

We have numerically calculated the values of $\alpha_c$,
$\alpha_N$ and $\eta/L$ for this model, as functions of the
coupling $h$ and the number of d.o.f. $g$ of the scalar
singlets. From these, we have calculated  the wall velocity
$v_{w}$ at $T=T_{N}$, using the results of section \ref{nume}.

\begin{figure}[htb]
\centering \epsfysize=7cm
\leavevmode \epsfbox{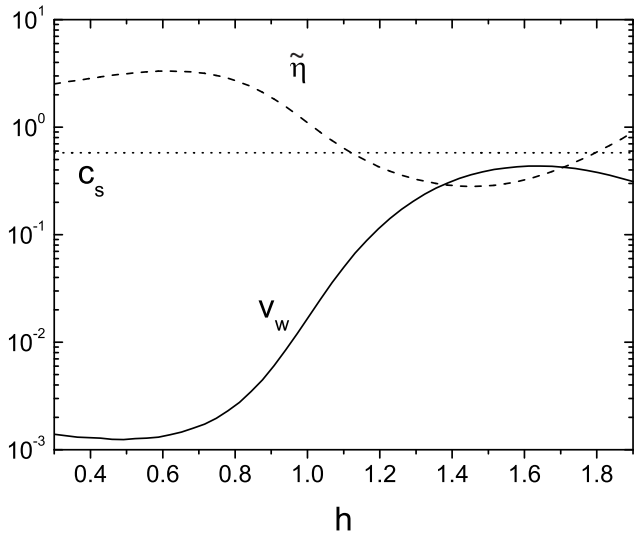}
\caption{The wall velocity and friction coefficient
as functions of the coupling $h$ for $g=2$ d.o.f.}
\label{figeta}
\end{figure}

In Fig. \ref{figeta} we have plotted the value of the
deflagration wall velocity as a function of $h$ for $g=2$.
Since the phase transition strengthens with $h$, one would
expect the wall velocity to be a monotonically increasing
function of $h$ (perhaps eventually bounded by $c_{s}$).
Indeed, roughly we have $v_{w}\left( T_{N}\right) \sim
\Delta\mathcal{F}\left( T_{N}\right) /\eta$, so $v_{w}$ is
dominated by the amount of supercooling and by the friction.
The friction coefficients $\eta_{\mathrm{th}}$ and
$\eta_{\mathrm{ir}}$ oscillate for $h$  $\sim 1$ due to the
$\log $ terms in Eqs. (\ref{etath}) and (\ref{etair}). This
causes a minimum and a maximum in the wall velocity. For large
$h$, the friction coefficient $\eta_{\mathrm{th}}$ dominates
since it goes like $h^{4}$, and the velocity finally decreases.
To illustrate this effect, we have plotted in Fig. \ref{figeta}
the value of the friction coefficient
$\tilde{\eta}=\tilde{\eta}_{\mathrm{th}}+\tilde{\eta}_{\mathrm{ir}}$.

\begin{figure}[hbt]
\centering \epsfysize=7cm
\leavevmode \epsfbox{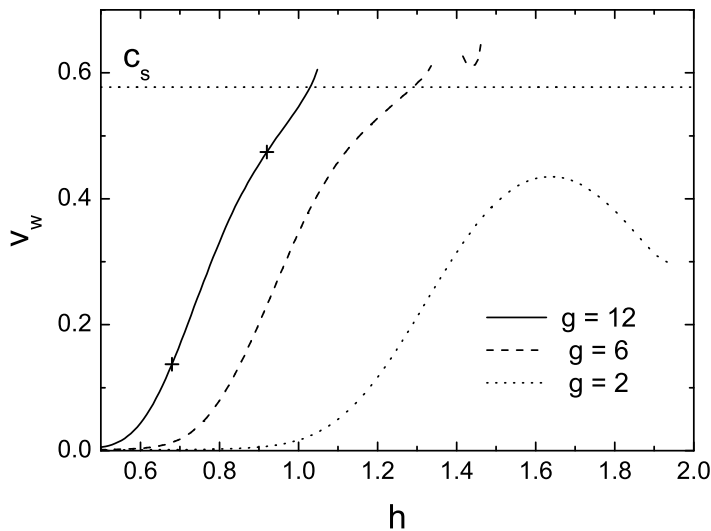}
\caption{The wall velocity as a function of $h$.}
\label{figmodel}
\end{figure}

Figure \ref{figmodel} shows the deflagration wall velocity  as a
function of the coupling $h$ for the cases $g=2$, $g=6$, and $g=12$.
The crosses in the $g=12$ curve indicate the points where $\phi
\left( T_{N}\right) /T_{N}\approx 1$ and $\phi \left( T_{N}\right)
/T_{N}\approx 2$. (These points were taken as a reference for the
variation of the parameters in the previous section.) For each  $g$,
the calculation was done up to a value $h=h_{\max }$ for which the
time required to get out of the supercooling stage becomes too long
for the numerical computation. This happens because the phase
transition becomes so strong that the barrier between minima persists
at low temperatures, and the nucleation temperature quickly falls to
zero. Indeed, near this endpoint, the free energy has a barrier
between minima already at $T=0$. For $h$ beyond $h=h_{\max }$, the
temperature $T_c$ also goes to $0$. The dependence on the coupling
will not be so strong if fermions are added to the model (see e.g.
Fig. 6 of Ref. \cite{ms08}).

Notice that the deflagration wall velocity can reach the speed
of sound $c_{s}$ only for large enough $g$ and the largest
values of $h$ in each curve, corresponding to really strong
phase transitions. Immediately after crossing the line of
$v_w=c_s$, the deflagration solution may disappear (cases $g=6$
and $g=12$ in the figure), and it may reappear (case $g=6$),
indicating a borderline case in parameter space. For the case
$m_{H}=125GeV$, we have considered values of $g_{b}$ in the
range $0-20$, varying $h$ up to $h_{\max }$ for each $g$, and
we have not found detonation solutions. This occurs because,
although the supercooling increases with $h$, the friction
increases as well. For lower Higgs masses, namely,
$m_{H}=100GeV$, detonation solutions appear only for $h\approx
h_{\max }.$

\begin{figure}[htb]
\centering \epsfysize=7cm
\leavevmode \epsfbox{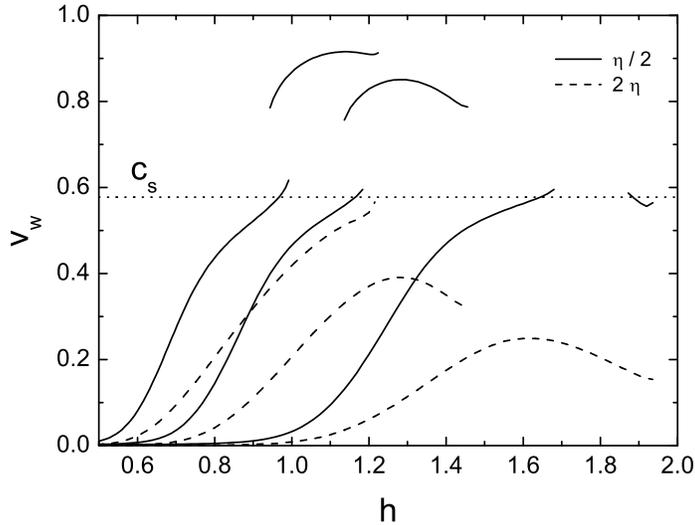}
\caption{Same as for Fig. \ref{figmodel}, but considering a larger and a
smaller friction parameter (from left to right, the curves
correspond to $g=12$, 6, and 2).}
\label{figetavar}
\end{figure}

Figure \ref{figexist} shows the values of $\alpha_c$ and
$\alpha_N$ corresponding to the curves of Fig. \ref{figmodel}.
Lower values of $h$ give weaker phase transitions and, hence,
lower values of $\alpha_c$ and $\alpha_N$. As $h$ becomes
large, the supercooling diverges before $\alpha_c$ reaches the
limiting value 1/3. Most points, though, are close to the
region where detonations do not exist. For these cases,
detonations would exist only for very low values of the
friction. On the other hand, for large $h$ the points move away
from the lowest curve in Fig. \ref{figexist}. For these points,
the supercooling is quite larger than the minimum that is
needed for the existence of detonations. Therefore, one expects
that this kind of solution would appear if the friction were a
little smaller. As we already mentioned, the error in the
calculation of the friction is a $\mathcal{O}(1)$ factor. Thus,
we also consider a friction which is a factor of 2 larger and a
factor of 2 smaller than the value given by Eqs.
(\ref{etath},\ref{etair}). The result is shown in Fig.
\ref{figetavar}. In the case $\eta/2$ we obtain detonations for
large values of $h$, if $g$ is large enough.

\section{Analytic approximations for the wall velocity} \label{aprox}

Analytic approximations for the wall velocity are very useful
for cosmological applications. Two approximations are commonly
used. In the case of deflagrations, if one ignores
hydrodynamics, the wall velocity can be obtained by equating
the pressure difference $\Delta p$ between phases (which pushes
the wall towards the high-temperature phase) to a friction
force per unit area, which is typically proportional to the
velocity, $f=\eta v_{w}$. Thus, one obtains the well known
result $ v_{w}=\Delta p(T)/\eta $, which has been widely used
in electroweak baryogenesis
\cite{dlhll92,lmt92,k92,mp95,mt97,m00,js01}. This velocity
depends on the amount of supercooling, since $\Delta p=0$ at
the critical temperature, and gives a good approximation for
nonrelativistic velocities. However, ignoring hydrodynamics
overestimates the value of $v_{w}$, since the released latent
heat accumulates in front of the wall, causing a slow-down of
the wall velocity. Nevertheless, if the friction is large
enough, this effect is negligible.

In the case of detonations, the Chapman-Jouguet condition,
$v_{-}=c_{s}$, leads to the simple formula
\begin{equation}
v_{J}\left( \alpha \right) =\frac{\sqrt{1/3}+\sqrt{\alpha ^{2}
+2\alpha /3}}{1+\alpha },  \label{detojug}
\end{equation}%
which is obtained by setting $v_{-}=1/\sqrt{3}$ in Eq.
(\ref{steinhardt}). This formula has been widely used for the
calculation of gravitational radiation \cite{kkt94,gw,eq07}.
However, as we have already mentioned, it is rather strange
that Eq. (\ref{detojug}) does not depend on microphysics.
Furthermore, it depends only on the ratio $ L/\tilde{\rho}
_{+}\left( T_{N}\right)$. This means that increasing the latent
heat will cause the same effect as decreasing $T_N$, which is
to increase the wall velocity. It is true that the pressure
difference goes like the product $L(T_c-T_N)$, but, on the
other hand, increasing $L$ will enlarge the effect of
hydrodynamics, which is to slow down the wall. As we shall see,
the velocity decreases if $L$ is increased with the amount of
supercooling held fixed \footnote{In a particular model,
increasing $L$ will give in general more supercooling. However,
such a relation cannot be derived from hydrodynamics alone, but
from calculating the nucleation rate.}. As can be seen in Fig.
\ref{figdeto}, according to our results, the Jouguet velocity
is a bad approximation to the detonation velocity (it
approximates well the non-physical solution).

In this section we shall derive analytical approximations both
for the detonation and the deflagration, by taking the
ultrarelativistic and nonrelativistic limits, respectively, of
the equations obtained in section \ref{eqs}. Before going on,
though, one further comment is worth. As we have seen, the
parameter $\varepsilon $ in $\alpha =\varepsilon /\tilde{\rho}
_{+}$ is not given by the latent heat $L,$ but by $ \varepsilon
=L/4$. This fact was already mentioned in earlier works on
gravitational waves \cite{kkt94}. However, the vacuum energy
density $ \varepsilon $ is sometimes confused with the latent
heat $L$. When analytical results are used for a particular
model, this amounts to considering a value of $\alpha $ four
times larger than the actual value, which leads to an
overestimation of the detonation velocity.

\subsection{The ultrarelativistic limit}

For $v_{w}\approx 1$, we can write $v_{w}=1-\delta $, with
$\delta \ll 1$. In this limit we must consider detonations, for
which $v_{w}=\left\vert v_{+}\right\vert $ and $\alpha
_{+}=\alpha _{N}\equiv \alpha $. Therefore, we have $\left\vert
v_{+}\right\vert =1-\delta $ and, from Eq. (\ref{steinhardt}),
$\left\vert v_{-}\right\vert =1-\delta \left( 1+3\alpha \right)
+\mathcal{O} \left( \delta ^{2}\right) $. To lowest order in
$\delta ,$ we have $\left\vert v_{+}\right\vert \gamma
_{+}=1/\sqrt{2\delta }$ and $\left\vert v_{-}\right\vert \gamma
_{-}=1/\sqrt{2\delta \left( 1+3\alpha \right) }$. Thus, the
last term in Eq (\ref{eqfric}) is $\sim 1/\sqrt{\delta }$. For
the first term we have, from Eq. (\ref{deltap}), $\left(
p_{+}-p_{-}\right) /\rho _{+}=-2\alpha +\mathcal{O}\left(
\delta \right) $. The second term is obtained from Eqs.
(\ref{st}) and (\ref{romeroma}). From Eq. (\ref{romeroma}) we
have $\rho _{-}/\rho _{+}=1+3\alpha +\mathcal{O}\left( \delta
\right) $, from which we obtain $T_{-}/T_{+}=\left(
a_{-}/a_{+}\right) ^{-1/4}\left( 1+3\alpha \right)
^{1/4}+\mathcal{O}\left( \delta \right) $, and
$s_{-}/s_{+}=\left( a_{-}/a_{+}\right) ^{1/4}\left( 1+3\alpha
\right) ^{3/4}+\mathcal{O}\left( \delta \right)$. Inserting
these results in Eq. (\ref{eqfric}) we obtain, to lowest
order\footnote{Keeping the next order in the equation is
straightforward, and leads to
\begin{equation} \sqrt{\delta}=\frac{\sqrt{\frac{32}{9}D^2-
\bar{\eta}^2(4q+3q^{1/2}+ 1-2/q)(q^{1/2}+1)}
-\frac{4\sqrt{2}}{3}D}{(\bar{\eta}/2)(4q+3q^{1/2}+ 1-2/q)},
\nonumber\end{equation} where $D$ is the denominator in Eq.
(\ref{ultra}) and $q=1+3\alpha$.},
\begin{eqnarray}
v_{w} &=&1-\delta , \notag \\
\sqrt{2\delta } &=&\frac{\left( 3/4\right) \left( \sqrt{1+3\alpha }+1\right)
\bar{\eta}}{r^{1/4}\left( 1+3\alpha \right)
^{5/4}-r^{-1/4}\left( 1+3\alpha \right)
^{3/4}},   \label{ultra}
\end{eqnarray}%
with $\alpha =\varepsilon /\left( a_{+}T_{N}^{4}\right) $,
$\bar{\eta}=\eta
/\left( a_{+}T_{N}^{4}\right) $, and $r\equiv a_{-}/a_{+}=1-3\alpha _{c}$, with $%
\alpha _{c}=\varepsilon /\left( a_{+}T_{c}^{4}\right) $. We remind
also that the correct relation between $\varepsilon$ and $L$ is
$\varepsilon =L/4$.

\begin{figure}[htb]
\centering \epsfysize=7cm
\leavevmode \epsfbox{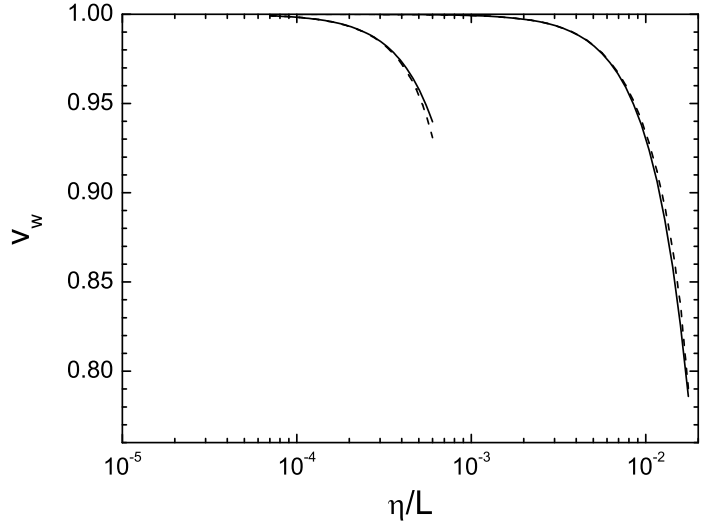}
\caption{The analytical approximation (\ref{ultra})  (dashed line)
together with the
numerical result (solid line), for the detonation solutions of Fig.
\ref{figdetodefla}, right.}
\label{figaproxdeto}
\end{figure}

Since the detonation velocity is usually large ($v_w\ge 0.8$),
the simple formula (\ref{ultra}) turns out to be an excellent
approximation in general. We have plotted this approximation in
Fig. \ref{figaproxdeto} together with the numerical result, for
the same detonation curves of the right panel of Fig.
\ref{figdetodefla}. For comparison, the Jouguet velocity lies
below 0.7 (and is constant for $\alpha$ fixed). Since the
parameter $\delta$ is bounded by $1-c_s\approx 0.4$, we expect
Eq. (\ref{ultra}) to give a good approximation in all the
possible range of the detonation velocity.

It is interesting to consider some limiting cases of Eq.
(\ref{ultra}). For $\eta \rightarrow 0$ with all the other
quantities held fixed, the wall velocity approaches the speed
of light. However, for any finite value of $\eta $, the
denominator diverges at $\alpha _{N}/\alpha
_{c}=a_{+}/a_{-}=1/\left( 1-3\alpha _{c}\right) $. Hence,
$\alpha _{N}$ has a lower bound larger than $\alpha _{c}$
(except in the limit $\alpha _{c}\rightarrow 0$). This
divergence reflects the fact that a certain amount of
supercooling is always needed for detonations to exist. As a
consequence, we cannot take the limit of small supercooling,
$\alpha _{N}/\alpha _{c}\approx 1.$

If we increase $L$ \emph{with fixed $\alpha_N$ and $\eta$}, it
can be easily seen from Eq. (\ref{ultra}) that $v_w$ decreases.
This is because increasing $L$ strengthens the effect of
hydrodynamics, which is to slow-down  the wall. Taking the
limit of large $L$ is not very useful. In the first place, $L$
is bounded by $(4/3)\tilde{\rho} _{+}\left( T_{c}\right) $ (we
remark that this is a general thermodynamical constraint
\cite{ms08}). Hence, this limit corresponds to taking $\alpha
_{c}\rightarrow 1/3$ (or $a_{-}\rightarrow 0$). However, this
is the limit of an extremely strong phase transition which, in
a realistic model will have also $T_{N}\rightarrow 0$ (or
$\alpha _{N}\rightarrow \infty $). Conversely, the limit of
large supercooling, $\alpha _{N}/\alpha _{c}\rightarrow \infty
$, should be taken together with $\alpha_{c}\rightarrow 1/3$.
It is not straightforward how to take this limit in Eq.
(\ref{ultra}), since the bag model does not provide a way to
calculate the nucleation temperature $T_{N}$.

On the other hand, we can consider the limit $L\rightarrow 0$.
In this case both $\alpha _{c}$ and $ \alpha _{N}\rightarrow
0$, (unless $T_{N}\ll T_{c}$, which will not be the case in a
realistic model if $L$ is small). Hence, Eq. (\ref{ultra})
yields $\sqrt{2\delta }\approx \left( \eta /\tilde{\rho}
_{+}\right) /\left( \alpha _{N}-\alpha _{c}\right) =\left(
4\eta /L\right) /\left( 1-T_{N}^{4}/T_{c}^{4}\right) $. From
Eqs. (\ref{eos}-\ref{leps}), we see that $L\left(
1-T_{N}^{4}/T_{c}^{4}\right) /4=p_{-}\left( T_{N}\right)
-p_{+}\left( T_{N}\right) $. Since in the ultrarelativistic
limit we have $1/\sqrt{2\delta }=v_{w}\gamma _{w}$, we obtain
$v_{w}\gamma _{w}=\Delta p(T_N)/\eta $. Thus, as expected, in
the limit $L\rightarrow 0$ we can neglect the effects of
hydrodynamics. The wall velocity becomes, according to Eq.
(\ref{ultra}), $v_{w}=1-\frac{1}{2}\left(\eta/ \Delta p(T_N)
\right) ^{2}$.

\subsection{The nonrelativistic limit}

In the nonrelativistic limit $v_{w}\ll 1$, we must consider
deflagrations. Thus, we have $\left\vert v_{-}\right\vert
=v_{w}$, and Eq. (\ref{steinhardt}) becomes $v_{+}=\left(
1-3\alpha _{+}\right) v_{-}+\mathcal{O}\left( v_{w}^{2}\right)
$. Therefore, according to Eq. (\ref{deltap}), we have $\left(
p_{+}-p_{-}\right) /\tilde{\rho} _{+}=\mathcal{O}\left(
v_{w}^{2}\right) $ and, by Eq. (\ref{romeroma}), $\rho
_{-}/\tilde{\rho} _{+}=1-3\alpha _{+}+ \mathcal{O}\left(
v_{w}^{2}\right) $. Thus, $T_{-}/T_{+}=\left(
a_{-}/a_{+}\right) ^{-1/4}\left( 1-3\alpha _{+}\right)
^{1/4}+\mathcal{O}\left( v_{w}^{2}\right) $ and
$s_{-}/s_{+}=\left( a_{-}/a_{+}\right) ^{1/4}\left( 1-3\alpha
_{+}\right) ^{3/4}+\mathcal{O}\left( v_w^2 \right)$. To lowest
order in $v_{w}$, Eq. (\ref{eqfric}) becomes
\begin{equation}
3\alpha _{+}\left( 2-3\alpha _{+}\right) \frac{\eta }{L}v_{w}=\left(
\frac{a_{-}}{a_{+}}
\right) ^{1/4}\left( 1-3\alpha _{+}\right) ^{3/4}-\left( \frac{
a_{-}}{a_{+}}\right) ^{-1/4}\left( 1-3\alpha _{+}\right)
^{1/4}+3\alpha _{+}.
\label{eqnr}
\end{equation}

In the case of deflagrations, we must also find $\alpha _{+}$
as a function of $\alpha _{N}$. In this approximation,
$v_{+}-v_{-}$ is given by $3\alpha _{+}v_{w}$. Hence, according
to Eq. (\ref{amndefla}), $\alpha _{N}-\alpha _{+}$ is of order
$v_{w}$, and so the left hand side becomes $\sqrt{3}\left(
\alpha _{N}-\alpha _{+}\right) /(4\alpha _{N})$. Thus, to
lowest order in $v_{w}$, Eq. (\ref{amndefla}) gives $ \alpha
_{+}=\alpha _{N}-4\sqrt{3}\alpha _{N}^{2}v_{w}$. Inserting this
in Eq. (\ref{eqnr}) and discarding higher order terms, we
obtain
\begin{multline}
v_{w}= \\
\frac{\left( 1-3\alpha \right) ^{\frac{3}{4}}\left( 3\alpha
-r^{-\frac{1}{4}}\left( 1-3\alpha \right)
^{\frac{1}{4}}+r^{\frac{1}{4}}\left( 1-3\alpha \right)
^{\frac{3}{4}}\right) }{\frac{3}{4}\bar{\eta}\left( 2-3\alpha \right)
\left( 1-3\alpha \right) ^{\frac{3}{4}}+3\sqrt{3}\alpha ^{2}\left[
r^{-\frac{1}{4}}+4\left( 1-3\alpha \right) ^{\frac{3}{4}}-3r^{\frac{1}{4}%
}\left( 1-3\alpha \right) ^{\frac{1}{2}}\right] }, \label{vwnrb}
\end{multline}
where $\alpha \equiv \alpha _{N} =\left( L/4\right) /\left(
a_{+}T_{N}^{4}\right)$, $\bar{\eta}=\eta /\left(
a_{+}T_{N}^{4}\right) $, and $r\equiv a_{-}/a_{+}=1-3\alpha
_{c}$, with $\alpha _{c}=\left( L/4\right) /\left(
a_{+}T_{c}^{4}\right) $.

The appearance of factors $\left( 1-3\alpha _{N}\right) ^{1/4}$
indicates the fact that the deflagration solution cannot exist
for $ \alpha _{N}$ arbitrarily large. For $\eta \rightarrow 0$,
only the first term in the denominator vanishes; the velocity
will not necessarily be large, nor the approximation
break-down. This reflects the fact that hydrodynamics slows
down the motion of the bubble wall. For $L\rightarrow 0$, i.e.,
$\alpha _{c},\alpha _{N}\rightarrow 0$, we obtain, similarly to
the detonation case, $v_{w}=\Delta p(T_{N})/\eta $, which
corresponds to neglecting hydrodynamics. It is interesting to
consider also, without neglecting hydrodynamics, the case in
which $\Delta p$ is small, i.e, the limit of small
supercooling.

\subsubsection{Small supercooling}

In the nonrelativistic limit, the pressure difference
$p_{+}-p_{-}$ is $\mathcal{O}\left( v_{w}^{2}\right) $. This
can be seen already from the nonrelativistic version of Eqs.
(\ref{eqlandau}),
\begin{equation}
w_{+}v_{+}=w_{-}v_{-},\ w_{+}v_{+}^{2}+p_{+}=w_{-}v_{-}^{2}+p_{-}.
\end{equation}
On the other hand, according to Eq. (\ref{eqmicro}), the
temperature difference $T_{+}-T_{-}$ is $\mathcal{O}\left( \eta
v_{w}\right) $. We shall now consider the case in which $\eta $
is not too small, so that the temperature difference must be
$T_{+}-T_{-}=\mathcal{O}\left( v_{w}\right) $. This means that,
to order $v_{w}$, the pressure difference due to supercooling,
$p_--p_+\sim T_c-T$, is canceled by the pressure difference due
to the temperature gradient, $p_--p_+\sim T_--T_+$. Hence, the
differences $T_{c}-T_{\pm }$ must be also $\mathcal{O} \left(
v_{w}\right) $. Thus, in this case we can expand the
thermodynamic quantities to first order in $T-T_{c}$.  In
particular, for the pressure functions we have $ p_{\pm }\left(
T\right) =p_{\pm }\left( T_{c}\right) +s_{\pm }\left(
T_{c}\right) \left( T-T_{c}\right) $, which can be used instead
of the equations of state (\ref{eos}). Thus, we can write
\begin{equation}
p_{+}\left( T_{+}\right) -p_{-}\left( T_{-}\right) =\left(
s_{+}-s_{-}\right) \left( T_{+}-T_{c}\right) +s_{-}\left( T_{+}-T_{-}\right)
,
\end{equation}%
where $s_{+}$ and $s_{-}$ are evaluated at $T=T_{c}$. Therefore
we have
\begin{equation}
T_{+}-T_{-}=\frac{\Delta s}{s_{-}}\left( T_{c}-T_{+}\right) +\mathcal{O}%
\left( v_{w}^{2}\right) .  \label{lineal}
\end{equation}%
To first order in $v_{w},$ we neglect the first term in Eq.
(\ref{eqmicro}) and, inserting Eq. (\ref{lineal}) in the second
term, we obtain
\begin{equation}
L\frac{T_{c}-T_{+}}{T_{c}}=\frac{s_{-}}{s_{+}}\eta v_{w},
\end{equation}%
where we have used $L=T_{c}\left( s_{+}-s_{-}\right) $ and
$v_+=(w_-/w_+)v_-$, with $|v_-|=v_w$.

If we had $T_{+}=T_{N},$ this would only give a correction of a factor $%
s_{+}/s_{-}$ to the usual equation $v_{w}=\Delta p\left(
T\right) /\eta $. However, an important contribution comes from
the fluid that is accumulated in front of the moving wall. For
$T_{+}$ and $T_{N}\approx T_{c},$ the relation (\ref{amndefla})
between the nucleation temperature $T_{N}$ and that of the
reheated fluid $T_{+}$ is given by
\begin{equation}
\frac{T_{+}-T_{N}}{T_{c}}=\frac{v_{w}L}{\sqrt{3}w_{+}},
\end{equation}%
where the enthalpy $w_{+}$ is evaluated at $T=T_{c}$. We finally
obtain
\begin{equation}
v_{w}=\frac{\left({w_{+}}/{w_{-}}\right)\left( T_{c}-T_{N}\right) /T_{c}}{\eta /L+L/(%
\sqrt{3}w_{-})}, \label{vwnr}
\end{equation}%
which agrees with the result of Ref. \cite{ikkl94}. Comparing
with Eq. (\ref{vwmicro}) and taking into account that $\Delta
\mathcal{F}\left( T\right) \approx L\left( T_{c}-T\right)/T_c
$, we may write
\begin{equation}
v_{w}=\frac{L\left( T_{c}-T_{N}\right)/T_{c} }{\eta _{\mathrm{eff}}},
\label{vweff}
\end{equation}%
where the effective friction coefficient
\begin{equation}
\eta _{\mathrm{eff}}=\frac{w_-}{w_+}\left(\eta + \frac{L^{2}}{\sqrt{3}w_{-}}\right)
\end{equation}
includes the effects of hydrodynamics. If friction dominates,
i.e., if $\eta /L\gg L/(\sqrt{3}w_{-})$, we obtain
$v_{w}=(w_-/w_+)L\left( T_{c}-T_{N}\right) /\eta $. For $L\ll
\rho _{+}$, we have $w_-\approx w_+$, and we recover Eq.
(\ref{vwmicro}). In the opposite limit, i.e., for $\eta /L\ll
L/(\sqrt{3}w_{-})$, we still obtain a finite velocity $
v_{w}=\left( 4\rho _{+}/L\right) \left( T_{c}-T_{N}\right) $.
Again, this means that the latent heat that is accumulated in
front of the bubble wall slows down the motion of the wall.

\begin{figure}[htb]
\centering \epsfysize=7cm
\leavevmode \epsfbox{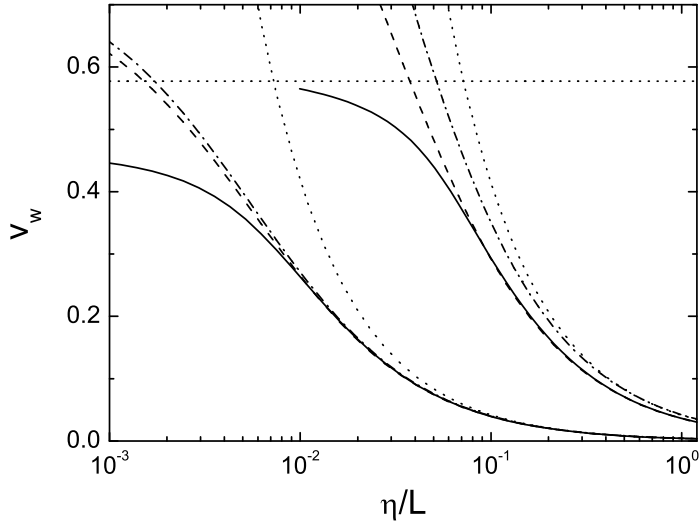}
\caption{The various analytical approximations for deflagrations. Solid
line: the numerical result. Dashed line: the nonrelativistic
approximation of Eq. (\ref{vwnrb}). Dashed-dotted line:
the approximation of Eq. (\ref{vwnr}) for small supercooling.
Dotted line: the usual approximation $v_w=\Delta p/\eta$.}
\label{figaproxdef}
\end{figure}

In Fig. \ref{figaproxdef} we plot the two approximations for
deflagrations, together with the numerical result and the usual
approximation which neglects hydrodynamics. As expected, the
numerical calculation gives smaller values of $v_{w}$ than the
analytical approximations, since the exact velocity has an upper
bound, namely, the speed of light. We see that in general the
nonrelativistic approximation is quite good up to $v_{w}\approx 0.4$,
and breaks down for $v_{w}\gtrsim c_{s}$. Notice that a useful, rough
approximation consists of assuming that $v_{w}$ is given by, say, Eq.
(\ref{vwnr}) up to $v_{w}=c_{s}$, and by $v_{w}=c_{s}$ when Eq.
(\ref{vwnr}) becomes supersonic.

\section{Conclusions} \label{conclu}

We have studied the steady state velocity of bubble  walls in
first-order cosmological phase transitions. We have numerically
investigated the whole velocity range $0<v_w<1$, and we have
found analytic approximations for the limits $v_w\ll 1$ and
$v_w\approx 1$.

Thus, we have considered both detonations and deflagrations. We
have derived a relatively simple set of equations for the wall
velocity and the fluid variables on both sides of the wall. To
obtain these equations we have used three ingredients, namely,
the continuity conditions for the fluid variables at the phase
transition front (and at the shock front in the case of
deflagrations), the introduction of a damping term in the
equation of motion for the order parameter (Higgs field), and
the bag equation of state. The parametrization of the friction
provides a model-independent equation which involves a friction
parameter $\eta$. We obtained this equation by generalizing a
procedure used in Ref. \cite{ikkl94} for the case of
nonrelativistic velocities and small supercooling. We have also
shown how to calculate the parameter $\eta$ in a specific
model.

We have solved numerically these equations for the wall
velocity as a function of the parameters $\alpha_c$, $\alpha_N$
and $\eta/L$ which depend essentially on the latent heat $L$,
the nucleation temperature $T_N$, and the friction coefficient
$\eta$. We have studied the regions in parameter space where
detonations and deflagrations are possible, and we have also
considered a specific model (the Standard Model with extra
singlet scalar fields) to investigate how a physical case moves
about through these regions.

We have found that supersonic velocities (either detonations or
deflagrations) are in general obtained only for very strong
phase transitions, with $\phi \left( T_{N}\right) /T_{N}> 2$,
occurring in the model for relatively large values of the
number of  scalar singlets $g$ and of their coupling $h$ to the
Higgs. Besides, the existence of detonations requires
relatively low values of the friction, and it may happen that
for such strong phase transitions neither detonations nor
deflagrations exist. In such a case, the steady state cannot be
reached, and the wall will accelerate until bubbles collide.
The case in which the wall velocity gets close to the speed of
light before bubble collision may have important implications
for gravitational wave generation. Interestingly, this
ultrarelativistic situation has been considered very recently
\cite{bm09} for the SM extension with singlet scalar fields,
finding that such ``runaway'' solutions exist for very strong
phase transitions.

In a more general extension of the SM, the singlets may have a
$\phi$-independent mass term, which will weaken the phase
transition. In that case, detonations may not appear at all.
The inclusion of fermions in the model will also weaken the
phase transition, but the friction coefficient will not have
the infrared contribution and will be smaller. In a forthcoming
paper \cite{ms09} we shall investigate several extensions of
the SM, including the MSSM and an extension with strongly
coupled fermions \cite{cmqw05}.

Finally, we have discussed the validity of commonly used
analytical approximations for the wall velocity, and we have
obtained nonrelativistic and ultrarelativistic approximations.
In particular, we have shown that the Jouguet velocity gives a
really bad approximation to detonations; the actual detonation
velocity is quite larger. We have found an alternative
approximation (\ref{ultra}) which, besides being more
realistic, is quite simple and gives an excellent fit to the
numerical result.

\section*{Acknowledgements}

This work was supported in part by Universidad Nacional de Mar del
Plata, Argentina, grants  EXA 365/07 and 425/08. The work by A.D.S.
was supported by CONICET through project PIP 5072. The work by A.M.
was supported by FONCyT grant PICT 33635.

\end{document}